\documentclass[a4paper,11pt]{article}
\usepackage{jheppub}
\usepackage{amsmath}
\usepackage{graphicx}
\usepackage{subfigure}
\usepackage{epstopdf}
\usepackage{epsfig}
\usepackage{amssymb}
\usepackage{bm}
\usepackage{bbm}
\newcommand{\beq}{\begin{equation}}
\newcommand{\eeq}{\end{equation}}
\newcommand{\ov}{\overline}
\newcommand{\pa}{\partial}

\usepackage{color}

\title{Higgs phenomenology in Type-I 2HDM \\ 
with $U(1)_H$ Higgs gauge symmetry}

\author[a]{P. Ko}
\author[b]{Yuji Omura}
\author[a]{Chaehyun Yu}
\affiliation[a]{School of Physics, KIAS, 85 Hoegiro, Seoul 130-722, Korea}
\affiliation[b]{Physik Department T30, Technische Universit\"{a}t M\"{u}nchen, \\
James-Franck-Stra$\beta$e, 85748 Garching, Germany}
\emailAdd{pko@kias.re.kr}
\emailAdd{yuji.omura@tum.de}
\emailAdd{chyu@kias.re.kr}

\abstract{
It is well known that generic two-Higgs-doublet models (2HDMs) suffer from potentially 
large Higgs-mediated flavor-changing neutral current (FCNC) problem, unless 
additional symmetries are imposed on the Higgs fields thereby respecting  
the Natural Flavor Conservation Criterion (NFC) by Glashow and Weinberg. 
A common way to respect the NFC is to impose $Z_2$ symmetry which is 
softly broken by a dim-2 operator. 
Another new way is to introduce local $U(1)_H$ Higgs flavor symmetry that distinguishes 
one Higgs doublet from the other.
In this paper, we consider the 
Higgs phenomenology in Type-I 2HDMs with the $U(1)_H$ symmetry with the simplest 
$U(1)_H$ assignments that the SM fermions are all neutral under $U(1)_H$,     
and we make detailed comparison with the ordinary Type-I 2HDM. 
After imposing various constraints such as vacuum stability and perturbativity as well 
as the electroweak precision observables and collider search bounds on charged Higgs 
boson, we find that the allowed Higgs signal strengths in our model are much broader 
than those in the ordinary Type-I 2HDM, 
because of newly introduced $U(1)_H$-charged singlet scalar and 
$U(1)_H$ gauge boson.   Still the ATLAS data on $gg\rightarrow h \rightarrow 
\gamma\gamma$ cannot be accommodated.  
Our model could be distinguished from the ordinary 2HDM with the $Z_2$ symmetry 
in a certain parameter region and some channels. 
If  the couplings of the new boson turn out to be  close to those in the SM, 
it would be essential to search for extra $U(1)_H$ gauge boson and/or one more 
neutral scalar boson to distinguish two models.
}

\begin{document}
%\pacs{}

\maketitle
\flushbottom

\section{Introduction}
The new boson discovered at the Large Hadron Collider (LHC) in the mass range  
125--126 GeV~\cite{Aad:2012tfa,Chatrchyan:2012ufa}  provides the missing link 
responsible for the origin of electroweak symmetry breaking and the masses of 
the Standard Model (SM) particles.   
Recent analyses for the spin and parity of this new boson at ATLAS and CMS exclude 
the hypothesis that this boson has different spin or parity from the SM Higgs boson 
by over 93\% C.L. or higher~\cite{Chatrchyan:2012jja,ATLAS-CONF-2012-169}.
Although there are controversial observations for the decay of the scalar boson,
such as the excess of the branching ratio for $h\to \gamma\gamma$ at ATLAS,  
the most updated values of couplings of this boson to the SM particles observed at 
the LHC indicate that this new boson is very close to the SM Higgs boson.
Then the next natural question on the scalar boson would be whether it is exactly the 
same as the SM Higgs boson, or one of Higgs bosons in Beyond SM with extended scalar 
sector.

One of the simplest extensions of the SM Higgs sector is the two-Higgs-doublet model 
(2HDM), where an extra Higgs $SU(2)_L$ doublet is added to the SM Higgs sector.
This extension may be motivated by many new physics models like the supersymmetric 
Standard Model, grand unified theories (GUTs), and so on.  
Many interesting physics issues have been studied in detail  within 2HDMs  
(see Ref.~\cite{review} for recent reviews).

However, the new scalars generally allow tree-level flavor-changing neutral currents 
(FCNCs) through the Yukawa couplings with SM fermions, and would be in conflict  
with observations that FCNC processes are highly suppressed in Nature, 
unless the scalars with flavor-changing tree-level couplings are heavy enough.\footnote{The FCNC problem mediated by the neutral Higgs boson may be
resolved in some specific models, where, for instance, the neutral Higgs couplings are naturally 
suppressed by the Cabibbo-Kobayashi-Maskawa matrix ($V_\textrm{CKM}$)~\cite{vckm}
or the Yukawa couplings are aligned in flavor space~\cite{alignment}.
}
 
One way to avoid this Higgs-mediated flavor problem is the so-called Natural Flavor
Conservation (NFC), where fermions of the same electric charges get their masses 
from one Higgs vacuum expectation value (VEV) \cite{Glashow}. One can assign new distinct charges  to the two 
Higgs doublets as well as to the SM fermions so that the NFC criterion can be 
achieved. Then the resulting Yukawa couplings involving the neutral scalars would not 
allow the tree-level  FCNCs mediated by neutral Higgs bosons.  

In most cases,  a softly broken discrete $Z_2$ symmetry is imposed in the 2HDMs 
\cite{Glashow}.  Two Higgs doublets, $H_1$ and $H_2$, have different $Z_2$ parity, 
and  only couplings following minimal flavor violation (MFV) are allowed. 
The 2HDMs with softly broken $Z_2$ symmetry {\it \`{a} la} the proposal of Glashow and 
Weinberg have been widely discussed in the literature, and  a lot of interesting signals 
can be predicted without serious conflicts with experiments involving FCNCs.
However, the predicted extra scalars in the 2HDMs  are strongly constrained by the 
collider search and the explicit $Z_2$ symmetry breaking terms tend to be required to 
shift the pseudoscalar mass.  Although this approach has been widely adopted in 
multi-Higgs doublet models, it is not clear what are the origins of the discrete $Z_2$ 
symmetry and its soft breaking. 

Recently the present authors proposed a new resolution of the Higgs-mediated FCNC 
problem in 2HDMs, by implementing the usual softly broken discrete $Z_2$ symmetry 
to spontaneously broken local $U(1)_H$ symmetry~\cite{Ko-2HDM}~\footnote{
See Ref.~\cite{thaler} for supersymmetric extension of the SM with extra 
gauge interactions including $U(1)_H$.}.  
Two Higgs doublets $H_1$ and $H_2$ have different $U(1)_H$ charges, and each SM 
fermion carries its own $U(1)_H$ charge in such a way that the phenomenologically 
viable Yukawa couplings are allowed without too excessive Higgs-mediated FCNC 
in a similar way to the usual 2HDMs with softly  broken $Z_2$ symmetry.   
The gauged $U(1)_H$ symmetry could realize such a large pseudo-scalar mass
by spontaneous breaking of  $U(1)_H$ gauge symmetry introducing a new SM singlet 
scalar $\Phi$ with nonzero $U(1)_H$ charge.
Then the local $U(1)_H$ symmetry is spontaneously broken into the softly broken 
$Z_2$ symmetry.  In other words, the 2HDMs with spontaneously 
broken local $U(1)_H$ symmetry could be the origin of the usual 2HDMs with softly  
broken $Z_2$ symmetry with the NFC criterion by 
Glashow and Weinberg.
In Ref.~\cite{Ko-2HDM}, the authors discussed in detail how to build new 2HDMs
with local $U(1)_H$ Higgs symmetry. In the type-I model, it is possible to construct
an anomaly-free model without extra chiral fermions by assigning appropriate $U(1)_H$ 
charges to the SM fermions and right-handed neutrino as in Table~I. 
It was also shown that the type-II 2HDM with local $U(1)_H$ symmetry could be 
interpreted as the effective  theory of the $E_6$ GUT model with leptophobic 
$Z^{'}$ boson \cite{London:1986dk,Rosner:1996eb}. 
These are new and amusing results, and the concept of local $U(1)_H$ Higgs gauge 
symmetry widely opens new possibilities for the multi-Higgs-doublet models.
 
The SM fermions are very often chiral under the $U(1)_H$ gauge symmetry proposed 
in Ref.~\cite{Ko-2HDM}, and the issues of anomaly cancellation and realistic Yukawa 
couplings have to be addressed carefully before one starts phenomenology.  
In general,  there appears gauge anomaly once extra gauge symmetry is added, 
so that extra chiral fermions are also required.  Also one may have to introduce new Higgs
doublets which are charged under new gauge groups, in order to write realistic Yukawa 
couplings.  When one discusses phenomenology in the extended SM with extra gauge 
symmetry,  one must consider all ingredients to make theory consistent,  even though 
some of them might be irrelevant at the electroweak energy scale.   
This procedure to include all ingredients to consist of phenomenological theory was 
emphasized in the chiral $U(1)^\prime$ models with flavored Higgs doublets, which could 
accommodate  the large deviation in the top quark forward-backward asymmetry at the 
Tevatron with the SM prediction~\cite{Ko-top1,Ko-top2,Ko-top3,Ko-B}.

Another new interpretation of the local $U(1)_H$ Higgs gauge symmetry proposed in 
Ref.~\cite{Ko-2HDM} is also possible.  Suppose there is a new chiral  local gauge 
symmetry in nature (to say, $U(1)_\chi$ for simplicity), under which some of the SM 
fermions are also charged.  Then it may be mandatory to extend the Higgs sector by 
introducing a new Higgs doublet  which is charged under the new chiral $U(1)_\chi$ 
gauge symmetry.  This is because in general one cannot write down the Yukawa 
couplings for  all the SM fermions without $U(1)_\chi$-charged Higgs doublets.  
The $U(1)_\chi$ charge of the Higgs doublet should match those of the SM chiral 
fermions in order to respect local $U(1)_\chi$ gauge symmetry.
There would be infinitely many possible choices for the $U(1)_\chi$ assignments which 
are also anomaly-free. However not all of them would be phenomenologically viable
because of the Higgs-mediated FCNC problem.  Only a subset of anomaly-free chiral 
$U(1)_\chi$ models with multi-Higgs-doublet models would satisfy the NFC criterion.
Our construction in Ref.~\cite{Ko-2HDM} can be regarded 
as finding new chiral $U(1)_\chi$ models which meet anomaly cancellation and the 
NFC {\it \`{a} la}  Glashow and Weinberg.

In this paper, we extend our previous work about the new 2HDMs with $U(1)_H$ Higgs 
symmetry ~\cite{Ko-2HDM}. In the previous work, we proposed $U(1)_H$ charge 
assignments  and full matter contents corresponding to each type of 2HDMs.
Since a SM-like Higgs boson was discovered at the LHC, it would be timely to discuss 
if our 2HDMs with local $U(1)_H$ symmetry would be consistent with the Higgs 
observation at the LHC.  After the discovery of the SM-like Higgs boson at the LHC,
a lot of works have been carried out in the context of the ordinary 2HDM of Type-I, 
Type-II,  Type-X, and Type-Y~\cite{Arhrib:2004ak,Arhrib:2003ph,Arhrib:2012ia,Mader:2012pm,Alves:2012ez,Altmannshofer:2012ar,Bai:2012ex,Bijnens:2013rd,Chiang:2013ixa,Krawczyk:2013gia,Barroso:2013zxa,Dery:2013aba,Coleppa:2013dya,Basso:2013wna,Ferreira:2013qua,Belanger:2013xza,Lopez-Val:2013yba,Chen:2013qda}.  
In this work, we will mainly concentrate on the simplest case,  the type-I 2HDM with 
$U(1)_H$ gauge symmetry, and compare our model with the ordinary type-I 2HDM. 
In the type-I 2HDM case, only one Higgs doublet couples to the SM fermions and the 
other Higgs doublet and singlet do not couple to them. 
In the type-I 2HDM with $U(1)_H$ symmetry, we can achieve anomaly-free models
without extra chiral fermions. Furthermore, constraints from flavor physics and
the collider experiments could be relaxed drastically (see Secs.~\ref{sec:EWPO} and \ref{sec:collider}).  

This paper is organized as follows. 
In Sec.~\ref{sec:potential}, we recapitulate the Type-I 2HDM with the spontaneous 
$U(1)_H$ Higgs gauge symmetry breaking including the general Higgs potential,  
and discuss the vacuum stability condition for the Higgs potential.
Then we derive the physical states of the Higgs fields and the masses of the $SU(2)$ 
gauge bosons in terms of the gauge coupling and Higgs VEVs and discuss the bounds on the physical masses of the charged Higgs and neutral Higgs bosons. Section~\ref{sec:EWPO} is devoted to the discussion of the constraints 
derived from electroweak precision observables (EWPOs),  and the comparison of our 
model with the usual Type-I 2HDM.  (The results obtained in Sec.~\ref{sec:EWPO} 
involves only gauge couplings of two Higgs  doublets and could be applied to and shared 
with other types of 2HDM~\cite{progress}.) 
Then we discuss phenomenology of Higgs bosons in our model at the LHC 
in Sec.~\ref{sec:collider}.  Conclusion of this paper is given in Sec.~\ref{sec:con}.
We present some useful formulas in Appendix.

\section{Type-I 2HDM with local $U(1)_H$ gauge symmetry}
\label{sec:potential}

\subsection{Generalities}
In 2HDMs, symmetry to distinguish the two $SU(2)_L$ Higgs doublets is required
in order to avoid tree-level FCNCs.   One usually assign $Z_2$ parities to two Higgs 
doublets and the SM fermion fields \cite{Glashow} to achieve the NFC by Glashow 
and Weinberg.   Depending on the charge assignment, one can obtain so-called 
Type-I 2HDM,  Type-II 2HDM, and etc.. 
Since the Yukawa couplings of the SM fermions are controlled by the $Z_2$ parities, 
the models allow the couplings respecting the hypothesis of MFV. 

In the usual 2HDMs with the softly broken $Z_2$ symmetry, there are extra physical 
scalar bosons:  one extra CP-even scalar ($H$),  one pseudoscalar ($A$), and one 
charged Higgs pair $(H^\pm)$.   The scalar masses are given by the Higgs VEVs 
and dimensionless couplings in the Higgs potential at the renormalizable level. 
Therefore we can expect that the mass scales of all extra scalar bosons are around 
the electroweak (EW) scale, 
like the SM-like Higgs boson observed at the LHC. 
However, the masses and couplings of the extra scalar bosons  are strongly constrained 
by the collider experiments and the EWPOs as well as the constraints from the flavor 
physics.  One has to introduce the $Z_2$ symmetry breaking term (soft breaking via 
dim-2 operators), which generates the pseudo scalar mass  ($m_A$), in order to 
consider the higher mass scales.

In Ref. \cite{Ko-2HDM}, the present authors proposed gauged $U(1)_H$ symmetry, 
which may be considered as  the origin of the $Z_2$ symmetry, and constructed a number 
of well-defined extensions of 2HDMs with only MFV.  In this case, the pseudo scalar mass 
$m_A$ is generated by spontaneous symmetry breaking of $U(1)_H$ via nonzero VEV
of a new $U(1)_H$-charged singlet scalar $\Phi$.
The Lagrangian for the two Higgs ($H_i~(i=1,2)$) and an extra $U(1)_H$-charged scalar 
$(\Phi)$ is
\beq
{\cal L}_H = \sum^2_{i=1} \left | \left( D^{SM}_\mu -i g_H q_{Hi} \Hat{Z}_{H \mu} \right) H_i 
\right |^2+  \left | \left( \pa_\mu -i g_H q_{\Phi} \Hat{Z}_{H \mu} \right) \Phi \right |^2
 - V_{\rm scalar} ( H_1, H_2 , \Phi ) + {\cal L}_{\rm Yukawa} ,
\eeq
where $D^{SM}_{\mu}$ is the covariant derivatives for $H_i$ under the SM-gauge groups. 
$g_H$ is the $U(1)_H$ gauge coupling, and $q_{Hi}$ and $q_{\Phi}$ are $U(1)_H$ 
charges of $H_i$'s and $\Phi$, respectively. $V_{\rm scalar}$ is the scalar 
potential for $H_i$ and $\Phi$ which breaks $U(1)_H$ and the EW symmetry.  
And $\Hat{Z}_{H\mu}$ is the $U(1)_H$ gauge boson in the interaction eigenstates.
Finally ${\cal L}_{\rm Yukawa}$ is the Yukawa interaction between the SM fermions 
and the two Higgs doublets, which would be  the same as the Yukawa interactions in 
Type-I, Type-II, etc..~\footnote{We ignore the kinetic mixing between $U(1)_H$ and 
$U(1)_Y$ for simplicity in this paper. }

This extension might suffer from tree-level deviation of the $\rho$ parameter due to the 
kinetic and mass mixings between the $U(1)_H$ gauge boson and $Z$ boson. 
Furthermore,  this extension would modify relevant collider signatures because of 
the additional Higgs doublet as well as the extra gauge boson $Z_H$ and the complex 
scalar $\Phi$.

\subsection{Type-I 2HDM with local $U(1)_H$ symmetry}  

There are many different ways to assign $U(1)_H$ charges to the SM fermions
to achieve the NFC in 2HDMs with local $U(1)_H$ gauge symmetry.  
The phenomenology will crucially depend on the $U(1)_H$ charge  assignments of 
the SM fermions. In general, the models will be anomalous,  even if $U(1)_H$ charge 
assignments are non-chiral, so that one has to achieve anomaly cancellation by adding 
new chiral fermions to the particle spectrum. 

%-----------------------------------------------------------------------

\begin{table}[th]
\begin{center}
\begin{tabular}{|c|cc|ccccc|}
\hline 
Type & $U_R$ & $D_R$ & $Q_L$ & $L$ & $E_R$ & $N_R$ & $H_2$ 
\\
\hline
$U(1)_H$ charge & $u$   & $d$   & $\frac{(u+d)}{2}$ & $\frac{-3 (u+d)}{2}$ & 
$-(2 u + d)$ & $-(u+2d)$ & $ q_{H_2}=\frac{(u-d)}{2}$
\\
\hline
$q_{H_1} \neq 0$ & $0$   & $0$    & $0$   & $0$    & $0$  & $0$  & $0$  
\\
$U(1)_{B-L}$ & $1/3$ & $1/3$  & $1/3$ & $-1$   & $-1$ & $-1$ & $0$ 
\\
$U(1)_R$  & $1$   & $-1$   & $0$   & $0$    & $-1$ & $1$  & $1$  
\\
$U(1)_Y$ & $2/3$ & $-1/3$ & $1/6$ & $-1/2$ & $-1$ & $0$  & $ 1/2$
 \\ \hline
\end{tabular}
\caption{
\label{table1}%
{
Charge assignments of an anomaly-free $U(1)_H$ in the Type-I 2HDM.
}
}
\end{center}
\end{table}

%--------------------------------------------------------------------

For the Type-I case, the present authors noticed that one can achieve an anomaly-free  
$U(1)_H$ assignment even without additional chiral fermions as in Table~\ref{table1}. 
Only $H_2$ couples with the SM fermions, and the $U(1)_H$ charges of $H_{1,2}$, 
$q_{H_1}$ and $ q_{H_2}$, should be different.  
Since the $U(1)_H$ charges of right-handed up- and down-type quarks ($u$ and $d$) 
in Table~\ref{table1} are arbitrary,  one can construct an infinite number of new models 
from the usual Type-I  2HDM by implementing the softly broken $Z_2$ symmetry to 
spontaneously broken local $U(1)_H$ gauge symmetry.
In the heavy $Z_H$ limit,  all the models with Type-I models 
with local $U(1)_H$ with arbitrary $u$ and $d$ will get reduced to the conventional 
Type-I 2HDM with softly broken $Z_2$ term (see $m_3^2$ term in Eq. (2) in the 
next subsection). 
In Table~\ref{table1}, we present four interesting $U(1)_H$ charge assignments:  the fermiophobic $U(1)_H$ with $u=d=0$, $U(1)_{B-L}$, $U(1)_R$, and $U(1)_Y$ cases.
  
\subsection{Scalar Potential} 
The scalar potential of general 2HDMs with $U(1)_H$ is completely fixed by local 
gauge invariance and renormalizability, and given by
%-----------------
\begin{eqnarray}
V_{\rm scalar}&=& \Hat{m}^2_1(|\Phi|^2) H^{\dagger}_1 H_1+ \Hat{m}^2_2(|\Phi|^2) H^{\dagger}_2 H_2  - \left ( m^2_{3} (\Phi) H^{\dagger}_1 H_2 
+h.c.  \right )  \nonumber \\
&&+ \frac{\lambda_1}{2} (H^{\dagger}_1 H_1)^2 + \frac{\lambda_2}{2} (H^{\dagger}_2 H_2)^2 
+ \lambda_{3} (H^{\dagger}_1 H_1)(H^{\dagger}_2 H_2)+ \lambda_{4} |H^{\dagger}_1 H_2|^2 \nonumber \\
&& +m^2_{\Phi} |\Phi|^2 + \lambda_{\Phi} |\Phi|^4.
\label{eq:potential}
\end{eqnarray}
%-----------------
$\Phi$ is a complex singlet scalar with $U(1)_H$ charge, $q_{\Phi}$, and contributes 
to the $U(1)_H$ symmetry breaking. 
$\Hat{m}^2_i(|\Phi|^2)$ $(i=1,2)$ and $m^2_{3} (\Phi)$ could be functions of $\Phi$:
$\Hat{m}^2_i(|\Phi|^2)=m_i^2+ \widetilde{\lambda}_i |\Phi|^2 $ at the renormalizable level.
$m^2_{3} (\Phi)$ is fixed by $q_{H_i}$ and $q_{\Phi}$, and $m^2_{3} 
(\langle \Phi \rangle)=0$  is satisfied at $\langle \Phi \rangle=0$:   $m^2_{3} (\Phi) = 
\mu \Phi^n$, where $n$ is defined as $n=(q_{H_1}-q_{H_2})/q_{\Phi}$.  A mass 
parameter $\mu$ can be regarded as real by suitable redefinition of  the phase of $\Phi$.   
Note that the $\lambda_5$ term ($\frac{1}{2} \lambda_5 [(H_1^\dagger H_2)^2+h.c.]$) 
in the usual 2HDMs with softly broken $Z_2$ symmetry does not appear in our models,  
because we impose the local $U(1)_H$ gauge symmetry instead of $Z_2$.  
In our model, the effective $\lambda_5$ term would be 
generated from the scalar exchange, after $U(1)_H$ symmetry breaking.
The effective $\lambda_5$ would contribute to the pseudoscalar mass, the vacuum 
stability and unitarity conditions like the ordinary 2HDMs. 
\footnote{The coupling $\lambda_5$ could also be generated by the dimension six 
operator $\lambda_5^\prime [(H_1^\dagger H_2)^2 \Phi^2 + h.c.] $.  
Then we have to keep  all the possible dimension-6 operators in the scalar potential 
in order to analyze the physical spectra which is a formidable task, and we would lose
the predictability.     
In this paper, we consider only the renormalizable lagrangian and just ignore higher 
dimensional operators for simplicity and predictability.}

Expanding the scalar fields around their vacua, 
\[
\langle H_i^T \rangle=(0,v_i/\sqrt{2}) , \ \ \ \langle \Phi \rangle=v_{\Phi}/\sqrt{2},
\]  
one can study the physical spectra in the scalar sector including  their masses and 
couplings.   The neutral scalars, $h_i,\chi_i,h_\Phi$, and $\chi_\Phi$,
and the charged Higgs, $\phi_i^+$,
in the interaction eigenstates are defined by
%-----------------
\begin{equation}
H_i = \begin{pmatrix} \phi^+_i  \\ 
\displaystyle
\frac{v_i}{\sqrt{2}} + \frac{1}{\sqrt{2}} (h_i + i \chi_i ) \end{pmatrix},
~~\Phi = \frac{1}{\sqrt{2}} (v_{\Phi} +h_{\Phi}+i \chi_{\Phi}).
\end{equation}
%-----------------
The scalar VEVs $v_i$ and $v_{\Phi}$ satisfy the stationary conditions 
(or vanishing tadpole conditions): 
%-----------------
\begin{eqnarray}
0&=&m^2_1 v_1 - m^{2}_{3}v_2 + \lambda_1 \frac{v^3_1}{2} 
+ \lambda_{3}\frac{v_1v_2^2}{2} +  \lambda_{4} \frac{v_1v_2^2}{2},  \\
0&=&m^2_2 v_2 - m^{2}_{3}v_1 + \lambda_2 \frac{v^3_2}{2} 
+ \lambda_{3}\frac{v_2v_1^2}{2} +  \lambda_{4} \frac{v_2v_1^2}{2},  \\
0&=& \frac{v_{\Phi}}{2} (\widetilde{\lambda}_1 v^2_1+\widetilde{\lambda}_2 v^2_2) - m'^2_3(v_{\Phi}) \frac{v_1v_2}{\sqrt{2}} +m^2_{\Phi} v_{\Phi}+\lambda_{\Phi} v^3_{\Phi},
\end{eqnarray}
with $m'^2_3(v_{\Phi})\equiv \pa_{\Phi}m^2_3(v_{\Phi}) $.

%-----------------

\subsection{Masses and Mixings of Scalar Bosons}
In 2HDMs with $U(1)_H$ and $\Phi$, there are three CP-even scalars, one pseudoscalar, 
and  one charged Higgs pair after $U(1)_H$ and EW symmetry breaking. 
There is also an additional massless scalar corresponding to $U(1)_H$ breaking, 
which  is eaten by the additional gauge boson of $U(1)_H$, called $Z_H$. 
Without $U(1)_H$-charged $\Phi$,  the two CP-odd scalars in $H_i$ could be eaten 
by the gauge bosons, so that we could discuss  the effective model with no massive 
pseudoscalar and $U(1)_H$ gauge boson \cite{Ko-2HDM,Lee:2013fda}.
One may consider a model with $Z_2$ Higgs symmetry instead of $U(1)_H$.
In this case, $\Phi$ should be a scalar to avoid a massless mode and three CP-even 
scalars  will appear after the symmetry breaking. Both cases will correspond to some 
limits of the 2HDM with $U(1)_H$ and $\Phi$.

\subsubsection{Charged Higgs $(H^\pm)$}
After the EW symmetry breaking, one Goldstone pair ($G^\pm$) 
and one massive charged Higgs pair ($H^\pm$) appear. 
The directions of Goldstone bosons are fixed by the Higgs VEVs: 
%-----------------
\beq
\begin{pmatrix} \phi_1^+ \\ \phi_2^+ \end{pmatrix} = \begin{pmatrix} \cos \beta \\ \sin \beta \end{pmatrix}G^+ + 
\begin{pmatrix} -\sin \beta \\ \cos \beta \end{pmatrix}H^+,
\label{mixinghp}%
\eeq
%-------------
where $(v_1,v_2)= (v \cos \beta, v \sin \beta)$ and $v=\sqrt{v_1^2+v_2^2}$. 
The squared mass of the charged Higgs boson $H^+$ is given by 
%-----------------
\begin{equation}
\label{eq:charged Higgs mass}
m^2_{H^{+}}= \frac{m_3^2}{\cos \beta \sin \beta} - \lambda_{4} \frac{ v^2}{2}.
\end{equation}
%-----------------
In the 2HDM without $\Phi$, $m_3^2$ is zero and 
$m^2_{H^{+}}$ is determined only by the second term with negative $\lambda_4$.
In the 2HDM with $\Phi$, $\lambda_4$ could be either negative or positive.

\subsubsection{Pseudoscalar boson $(A)$}\label{sec:pseudoscalar}
In 2HDMs with discrete $Z_2$ symmetry, one CP-odd mode is eaten by the $Z$ boson 
and the other  becomes massive. In the 2HDM with a complex scalar, $\Phi$, there is an 
additional CP-odd mode  and two Goldstone bosons ($G_{1,2}$) appear after the EW 
and $U(1)_H$ symmetry breaking.
$m_3^2(\Phi)$ plays a crucial role in the mass of $A$, $m_A$. $m_3^2(\Phi)$ is
$m_3^2(\Phi)=\mu \Phi$ or $\mu \Phi^2$ in the renormalizable potential
depending on the definition of $q_{\Phi}=(q_{H_1}-q_{H_2})$ or $(q_{H_1}-q_{H_2})/2$.

The directions of $G_{1,2}$ and $A$ are defined as  
\begin{eqnarray}
\begin{pmatrix} \chi_{\Phi}\\ \chi_1 \\ \chi_2  \end{pmatrix}&=& \begin{pmatrix} 0 \\  \cos \beta  \\   \sin \beta  \end{pmatrix} G_1+\frac{v_{\Phi}}{\sqrt{v^2_{\Phi} + (n v \cos \beta \sin \beta)^2}}
\begin{pmatrix} 1 \\ \frac{n v}{v_{\Phi}} \cos \beta \sin^2 \beta \\  - \frac{n v}{v_{\Phi}} \cos^2 \beta \sin \beta  \end{pmatrix} G_2 \nonumber  \\
&+&\frac{v_{\Phi}}{\sqrt{v^2_{\Phi} +  (n v \cos \beta \sin \beta)^2}}
\begin{pmatrix} \frac{n v}{v_{\Phi}} \cos \beta \sin \beta \\ - \sin \beta \\   \cos \beta  \end{pmatrix} A ~ .
\label{mixingpseudo}%
\end{eqnarray}
The squared pseudoscalar mass $m_A^2$ is given by
\beq
m^2_{A}= \frac{m_3^2}{\cos \beta \sin \beta} \left ( 1+ \frac{n^2v^2}{v^2_{\Phi}} \cos^2 \beta \sin^2 \beta \right ),
\eeq
where $n=1$ or $2$ depending on $m_3^2(\Phi)$.
$G_1$ corresponds to the Goldstone boson in the ordinary 2HDMs and could be eaten 
by the $Z$ boson.  In the limit, $v_{\Phi} \to \infty$, $\chi_{\Phi}$ is $G_2$ and eaten by 
$Z_H$.   Also the direction of $A$ and $m^2_{A}$ become the same as in the ordinary 
2HDMs.  In the 2HDM with local $U(1)_H$ symmetry but without $\Phi$, $A$ does not 
exist, so that it could corresponds to the limit, $m_A \to \infty$ and $v_{\Phi} \to 0$. 
In the following section,  we discuss our 2HDMs assuming $m_3^2(\Phi)=\mu \Phi$ and 
$q_{\Phi}=(q_{H_1}-q_{H_2})$.

\subsubsection{CP-even scalar bosons $(h,H, \tilde{h})$}\label{sec:CP-even}
After the EW and $U(1)_H$ symmetry breaking, three massive CP-even scalars 
appear and they generally mix with each other as follows: 
\beq
\begin{pmatrix} h_{\Phi}\\ h_1 \\ h_2  \end{pmatrix} =   \begin{pmatrix} 1 & 0 & 0 \\ 0 & \cos \alpha & -\sin \alpha  \\ 0& \sin \alpha & \cos \alpha \end{pmatrix}  \begin{pmatrix} \cos \alpha_{1}& 0  & -\sin \alpha_{1}\\ 0 & 1 & 0   \\ \sin \alpha_{1}& 0 & \cos \alpha_{1} \end{pmatrix}\begin{pmatrix} \cos \alpha_{2} & -\sin \alpha_{2} & 0 \\ \sin \alpha_{2} & \cos \alpha_{2} & 0  \\ 0 & 0 & 1  \end{pmatrix} \begin{pmatrix} \widetilde{h} \\ H \\ h  \end{pmatrix},
\label{mixingneutral}%
\eeq
where $\alpha$ corresponds to the mixing angle between two neutral scalars 
in the ordinary 2HDM and $\alpha_{1,2}$ are additional mixing angles that newly appear 
in our model with local $U(1)_H$ and a singlet scalar $\Phi$.
The mixing is given by the mass matrix which is introduced in Appendix 
\ref{sec:Mass Matrix of CP-even scalars}.   In the limit of $\alpha_{1,2}\to 0$  
one can interpret $h_{\Phi}$ as the field in the mass basis and $h_\Phi$ does not mix 
with $h_{1,2}$.  Throughout this paper, we assume that $h$ is the SM-like scalar 
boson with its mass ($m_h$) being fixed around $126$ GeV.

\subsection{Gauge bosons}
In 2HDMs with local $U(1)_H$ Higgs symmetry,  at least one of the Higgs doublets 
$H_{i=1,2}$ should be charged under $U(1)_H$. Therefore tree-level mass mixing 
between $Z$ and $Z_H$ would appear after spontaneous breaking of the EW
and $U(1)_H$ symmetries.
Let us describe the mass matrix of $Z$ and $Z_H$ as
\beq
\begin{pmatrix}  \Hat{M}^2_{Z} & \Delta M_{ZZ_H}^2 \\  \Delta M_{ZZ_H}^2 & \Hat{M}^2_{Z_H } \end{pmatrix}.
\eeq
$\Hat{M}^2_{Z}$ and $\Hat{M}^2_{Z_H}$ are
\beq
\Hat{M}^2_{Z}= \frac{g^2+g'^2}{4}v^2 =\frac{g_Z^2}{4} v^2,~ \Hat{M}^2_{Z_H }= g_H^2 \left \{ \sum^2_{i=1} (q_{H_i} v_i)^2+ q^2_{\Phi} v^2_{\Phi} \right \},
\eeq 
and the mass mixing term between $Z$ and $Z_H$ is
\beq
\Delta M_{ZZ_H}^2= -\frac{\Hat{M}_{Z}}{v} g_H\sum^2_{i=1}  q_{H_i}  v^2_i. 
\eeq
Here $g, g^\prime$ and $g_H$ are the gauge couplings of $U(1)_Y$, $SU(2)_L$, and 
$U(1)_H$ gauge interactions, respectively. And $q_{H_i}$ and $q_\Phi$ are the $U(1)_H$ 
charges of the Higgs doublet $H_i$'s and the singlet scalar  $\Phi$, respectively. 
Some examples of the charge assignments within Type-I 2HDM are shown in Table \ref{table1}.  $U(1)_H$ charge assignments for other types of 2HDMs can be found in 
Ref.~\cite{Ko-2HDM}. 

The tree-level masses in the mass eigenstates are given by
\begin{eqnarray} \label{eq:Z mass}
 M_{Z0}^2 &=& \frac{1}{2} \left \{ \Hat{M}^2_{Z_H}+\Hat{M}^2_{Z} -\sqrt{(\Hat{M}^2_{Z_H}-\Hat{M}^2_{Z} )^2 +4 \Delta M_{ZZ_H}^4}   \right \}, \\
 M_{Z_H0}^2 &=& \frac{1}{2} \left \{ \Hat{M}^2_{Z}+\Hat{M}^2_{Z_H} +\sqrt{(\Hat{M}^2_{Z_H}-\Hat{M}^2_{Z} )^2 +4 \Delta M_{ZZ_H}^4}   \right \} . 
\end{eqnarray}
Then the mixing between $Z$ and $Z_H$ is described by the mixing angle $\xi$, 
which is defined as
\beq
\label{eq:mixing}
\tan 2 \xi = \frac{2 \Delta M_{ZZ_H}^2}{\Hat{M}^2_{Z_H}-\Hat{M}^2_{Z}}. 
\eeq
Note that we omit the symbol ``0" for the physical (renormalized) masses for the gauge bosons.  
The extra gauge boson couples with the SM fermions through the mixing even if the SM fermions are not charged
under $U(1)_H$. Furthermore, this mixing modifies the coupling of the $Z$ boson with the fermions,  which has been well-investigated at the LEP experiments.
The $Z$ boson mass is also deviated from the SM prediction according 
to Eq.~(\ref{eq:Z mass}) and the allowed size of the deviation is 
evaluated by the $\rho$ parameter. Our 2HDMs are strongly constrained 
not only by the $Z_H$ search in the experiments  but also by the EWPOs, 
as we will see in the next section.

\section{Vacuum Stability Condition and Various Constraints\label{sec:EWPO}}
\subsection{Vacuum stability condition and perturbative unitarity bounds 
\label{sec:vacuum stability}}
There are many theoretical and experimental constraints on our model.
First we consider theoretical bounds on Higgs self couplings from vacuum stability 
condition and perturbative unitarity.

In order to break the $U(1)_H$ and EW symmetry,  the potential (\ref{eq:potential}) should 
have a stable vacuum with nonzero VEVs, namely the scalar potential is bounded 
from below.   We impose the vacuum stability bounds, which require that  
the dimensionless couplings $\lambda_{1,2,3,4}$ are to satisfy the following conditions: 
\beq\label{eq:vacuum-condition}
\lambda_1 >0 ,~ \lambda_2>0 ,~ \lambda_3  > - \sqrt{\lambda_1 \lambda_2} ,
~ \lambda_3 +\lambda_4  > - \sqrt{\lambda_1 \lambda_2},
\eeq
in the $\langle \Phi \rangle = 0 $ direction.
They correspond to the ones in the usual 2HDMs without $\lambda_5$.
Following the conditions and  Eq. (\ref{eq:relation-2HDM}) in 
Appendix \ref{sec:Mass Matrix of CP-even scalars},  the masses of scalars satisfy 
\beq\label{eq:relation-2HDM-0}
m_h^2 +m_H^2-m_A^2 > 0.
\eeq
In the ordinary 2HDMs with softly broken $Z_2$ symmetry,
sizable $\lambda_5$ is allowed and 
the conditions (\ref{eq:vacuum-condition}) and (\ref{eq:relation-2HDM-0})
should be modified by the replacements, $m_{H^+}^2 \to m_{H^+}^2 +\lambda_5 v^2$
$m_A^2 \to m_A^2 +\lambda_5 v^2$ and $\lambda_4 \to \lambda_4 - | \lambda_5 |$ 
in Eqs.~(\ref{eq:charged Higgs mass}),
 (\ref{eq:vacuum-condition}), and (\ref{eq:relation-2HDM-0}).

In the $\langle \Phi \rangle \neq 0$ direction, the vacuum-stability conditions for $\lambda_{\Phi}$, $\widetilde{\lambda_1}$ and $\widetilde{\lambda_2}$ are
\begin{eqnarray}
\lambda_{\Phi}>0,~\lambda_1 > \frac{\widetilde{\lambda_1}^2}{\lambda_{\Phi}},~ \lambda_2&>&\frac{\widetilde{\lambda_2}^2}{\lambda_{\Phi}},  ~\lambda_3 -  \frac{\widetilde{\lambda_1}\widetilde{\lambda_2}}{\lambda_{\Phi}} > - \sqrt{ \left( \lambda_1-\frac{\widetilde{\lambda_1}^2}{\lambda_{\Phi}} \right) \left( \lambda_2-\frac{\widetilde{\lambda_2}^2}{\lambda_{\Phi}} \right)}, \nonumber \\
\lambda_3+\lambda_4  -  \frac{\widetilde{\lambda_1}\widetilde{\lambda_2}}{\lambda_{\Phi}}& >& - \sqrt{ \left( \lambda_1-\frac{\widetilde{\lambda_1}^2}{\lambda_{\Phi}} \right) \left( \lambda_2-\frac{\widetilde{\lambda_2}^2}{\lambda_{\Phi}} \right)},
\end{eqnarray}
where the directions of $H_1$ and $H_2$ fields in the last four conditions 
are the same as those of $H_1$ and $H_2$ fields in Eq.~(\ref{eq:vacuum-condition}).

We also impose the perturbativity bounds $\lambda_i \le 4 \pi$ on the quartic Higgs 
couplings and the tree-level unitarity conditions whose expressions are given
in Ref.~\cite{unitarity1,unitarity2,unitarity3}.  These will make theoretical constraints on the quartic couplings
in the scalar potential (2).

\subsection{Constraints from various  experiments\label{sec:constraints from the experiments}}

The charged Higgs boson mass is constrained by the LEP experiments. It depends
on the decay channel of the charged Higgs boson, and we take the model-independent 
bound $m_{h^+} \gtrsim  80$ GeV~\cite{lephp} in this work.
We also impose  a recent bound on the charged Higgs and $\tan\beta$ coming from 
the top quark decay from the LHC experiments~\cite{topbound1,topbound2,topbound3}.
We note that the flavor bound which mainly comes from the $b\to s \gamma$ experiments 
is $\tan \beta \gtrsim 1$ in the type-I 2HDM \cite{Hermann:2012fc}. 

Recently the BABAR Collaboration reported about $3.4\sigma$ deviation
from the SM prediction in the $B\to D^{(\ast)}\tau\nu$ decays~\cite{dtaunu}.
This deviation cannot be accommodated with the ordinary 2HDM with MFV 
in the Yukawa sector. It turned out that 2HDMs which violate
MFV might account for the discrepancy.  The chiral $U(1)^\prime$ model
with flavored Higgs doublets which slightly breaks the NFC criteria
in the right-handed up-type quark sector~\cite{Ko-B} is one of such examples.
Since the 2HDMs with $U(1)_H$ hold the MFV hypothesis, they cannot
be accommodated with the deviation in $B\to D^{(\ast)}\tau\nu$.
In this work, we do not consider these experiments seriously since 
the experimental results are not well settled down.
In the future, if this deviation would be confirmed at Belle or Belle II,
it might exclude our 2HDMs as well as the ordinary 2HDMs.

EWPOs in the LEP experiments 
which are usually parametrized by Peskin-Takeuchi parameters $S$, $T$, 
and $U$~\cite{peskin} 
provides strong bounds on the parameters in the Higgs potential.
If new physics has no direct couplings to the SM fermions,  their effects at the LEP 
energy scale would appear only through the self energies of $SU(2)_L$ gauge bosons. 
This is the case of the usual Type-I 2HDM. 
However, in our model there exists a new $U(1)_H$ gauge boson, which may couple 
to the SM fermions. In this case, one must consider all observables at the $Z$ pole 
at the one-loop level instead of $S$, $T$, and $U$~\cite{Erler:2009jh}. 
However if the new gauge boson is decoupled from the EW scale physics,
$S$, $T$, and $U$ will provide well-defined constraints on the 2HDMs
with $U(1)_H$. We will discuss this bound in a few next subsections.

\subsection{Tree-level $\rho$ parameter}
If the Higgs doublets are charged under the extra gauge symmetry,  the extra symmetry 
would also be broken along with EW symmetry breaking. 
Then there appears the mass mixing 
between the $Z$ boson and the extra massive gauge boson.
In the 2HDMs with $U(1)_H$, the mixing between $Z$ and $Z_H$ is generated as 
in Eq.~(\ref{eq:mixing}). This mass mixing could allow the $Z$ boson mass to deviate
significantly from the SM prediction, and thus  will strongly be constrained by  the 
$\rho$ parameter, which the SM predicts to be one at the tree level. 

Assuming $\xi \ll 1$, the tree-level $\rho$ parameter is described as
\beq
\rho =1+ \frac{ \Delta M_{ZZ_H}^2}{M^2_{Z0}} \xi +O(\xi^2).
\eeq
The mixing also changes the $Z$ boson couplings with the SM fermions and the factor is estimated as $1-\xi^2/2$.

The bounds on the tree-level mixing have been discussed 
in Refs.~\cite{Cheng-Wei,Babu:1996vt,Babu:1997st}. 
As we will see in Fig. \ref{fig:Bound-STU} (a), we can derive the bounds 
on $g_H$, $\tan \beta$, and $M_{Z_H}$ in the case with 
$(q_{H_1},q_{H_2})=(1,0)$, when we require that the tree-level contributions 
to the $\rho$ parameter and 
the decay width of the $Z$ boson, 
which are functions of the $Z$-boson couplings,
are within the error of 
the SM predictions: $\rho=1.01051\pm 0.00011$ 
and $\Gamma_Z=2.4961 \pm 0.0010$ GeV \cite{EWPO-PDG}. 
The tree-level deviations may also affect the $S$, $T$, and $U$ parameters, but
they actually become negligible because of the requirement for the stringent 
bound from $Z'$ search at the LHC, as we discuss in the next section.

\subsection{Bound from $Z'$ search in the collider experiments}
\label{sec:Zprime search}
Extra neutral gauge bosons are strongly constrained by $Z'$ searches at high energy 
colliders. 
In our models, $Z_H$ can couple with the SM fermions through the $Z$-$Z_H$ mixing, 
even if we choose  the charged assignment that the SM fermions are not charged under  
$U(1)_H$.
 
If $Z_H$ couples with leptons, especially electron and muon,  $Z_H$ would be produced  
easily at LEP and the coupling and mass of $Z_H$ are strongly constrained by the 
experimental results, which are consistent with the SM prediction with very high accuracy.
If $Z_H$ is heavier than the center-of-mass energy of LEP ($209$ GeV),
we could derive the bound on the effective coupling of $Z_H$ 
\cite{Carena:2004xs,LEP:2003aa,Alcaraz:2006mx}. 
The lower bound on $M_{Z_H}/g_H$ would be $O(10)$ TeV \cite{LEP:2003aa,Alcaraz:2006mx}. 
If $Z_H$ is lighter than $209$ GeV,
the upper bound of $Z_H$ coupling would be $O(10^{-2})$ to avoid conflicts 
with the data of $e^+ e^- \to f^- f^+$ $(f=e, \mu)$ \cite{EWPO-PDG,LEP:2003aa,Alcaraz:2006mx}.

Furthermore, there will be strong bounds from hadron colliders,
if quarks are charged under $U(1)_H$. 
The upper bounds on the $Z_H$ production at the Tevatron and LHC 
are investigated in the processes, $p p(\overline{p}) \to Z_H X \to f \overline{f} X$ \cite{EWPO-PDG,Carena:2004xs,LHCZprime-ATLAS,LHCZprime-CMS}, and the stringent bound requires $O(10^{-3})$ times
smaller couplings than the $Z$-boson couplings 
for $M_{Z'} \leq 1$ TeV~\cite{LHCZprime-CMS}.

We could avoid these strong constraints,
in the case that all particles except for one Higgs doublet are not charged under $U(1)_H$.
Actually the model in the first row of Table~\ref{table1} is this case.
$Z_H$ couples with the SM fermions only through the $Z$-$Z_H$ mixing, so that the mixing should be
sufficiently small. In the following sections, we focus on the fermiophobic $U(1)_H$ charge assignment
and require the (conservative) bound $\sin \xi \lesssim 10^{-3}$, 
according to Ref.~\cite{LHCZprime-CMS}. The small mixing especially contributes to the $T$ parameter as $\alpha T \sim \rho-1$, but it will not affect our results.

In the 2HDM with $U(1)_H$, $Z_H$ can decay to $Z$ and scalars, so that
the strong bound, $\sin \xi \lesssim 10^{-3}$, will be relaxed 
if the branching ratio of the $Z_H$ decay into $Z$ and scalars
is almost one.
In the following sections, we study the region with $M_{Z_H} \leq 1$TeV , and
the additional branching ratio is at most $0.1$ in that region.
If we assume that there are extra particles charged under $U(1)_H$ and $Z_H$ 
mainly decays to the extra particles,
the larger value for $\sin \xi$ could be allowed.
We note that the constraint from the $Z^\prime$ search
in the dijet production at the LHC can easily be avoided by the bound
on the mixing angle $\xi$.

In the region of $M_{Z_H} >1$ TeV, the constraints from the $Z'$ search are relaxed 
and the constraint on $g_H \cos \beta$ from the $\rho$ parameter and $\Gamma_Z$
becomes stronger as we will see in Fig.~\ref{fig:Bound-STU} (a).

\subsection{$S$, $T$, and $U$ parameters at the one-loop level}
Here, we introduce $S$, $T$, and $U$ parameters in the 2HDMs with the $U(1)_H$
gauge boson and $\Phi$ at the one-loop level. 
They involve only gauge interactions of scalars, so that the results could be applied 
to other types of 2HDMs~\cite{progress}.
The EWPOs in 2HDMs with extra scalars have been calculated in Refs.\cite{EWPO-2HDMwScalars1,EWPO-2HDMwScalars2}.

In order to calculate the $S$, $T$, and $U$ parameters, we define mass
eigenstates $\{H_l^+\}$, $\{H_l\}$, and $\{A_l\}$
of Higgs bosons in terms of mixing angles
$\beta$, $\alpha$, and $\alpha_{1,2}$,
\begin{equation}
\phi^+_i=c^{H^+_l}_{\phi_i} H^+_l,~h_i=c^{H_l}_{h_i} H_l,
~\chi_i=c^{A_l}_{\chi_i} A_l,
\end{equation}
where $\{ H^+_l \}=(G^+,H^+ )$, 
$\{ H_l \}=(\widetilde{h},H,h)$, $\{ A_l \} = (G_1,G_2,A)$.
The masses of Goldstone bosons are given by
$m_{G^+}=M_W$, $m_{G_1}=M_{Z}$ and $m_{G_2}=M_{Z_H}$ 
in the Feynman gauge.
$c^{H^+_l}_{\phi_i}$, $c^{\Hat{h}_l}_{h_i}$, and $c^{A_l}_{\chi_i}$ satisfy
\begin{eqnarray}
\sum_l c^{H^+_l}_{\phi_i} c^{H^+_l}_{\phi_j} = \delta_{ij},~&&\sum_l c^{H_l}_{h_i}c^{H_l}_{h_j}= \delta_{ij},~\sum_l c^{A_l}_{\chi_i}c^{A_l}_{\chi_j}= \delta_{ij},  \\
\sum_i c^{H^+_l}_{\phi_i} c^{H^+_m}_{\phi_i} = \delta_{lm},~&&\sum_i c^{H_l}_{h_i}c^{H_m}_{h_i}+c^{H_l}_{h_{\Phi}}c^{H_m}_{h_{\Phi}}= \delta_{lm},~\sum_i c^{A_l}_{\chi_i}c^{A_m}_{\chi_i}+c^{A_l}_{\chi_{\Phi}}c^{A_m}_{\chi_{\Phi}}= \delta_{lm}.
\end{eqnarray}
Each mixing angle is given in Eqs.~(\ref{mixinghp}), (\ref{mixingpseudo}),
and (\ref{mixingneutral}).

Let us discuss the constraints on the loop corrections to the EWPOs in terms of the  
$S$, $T$, and $U$ parameters defined as 
\cite{EWPO-PDG}
\begin{eqnarray}
\alpha(M^2_Z)T&=&\alpha(M^2_Z)T_{\rm 2HDM}+ \frac{ \Delta \Pi_{WW}(0)}{M^2_W} -\frac{ \Delta \Pi_{ZZ}(0)}{M^2_Z},  \\
\frac{\alpha(M^2_Z)}{4s^2_Wc^2_W}S&=&\frac{\alpha(M^2_Z)}{4s^2_Wc^2_W}S_{\rm 2HDM}+ \frac{ \Delta \Pi_{ZZ}(M^2_Z)- \Delta \Pi_{ZZ}(0)}{M^2_Z},  \\
\frac{\alpha(M^2_Z)}{4s^2_W}(S+U)&=&\frac{\alpha(M^2_Z)}{4s^2_W}(S_{\rm 2HDM}+U_{\rm 2HDM})+\frac{\Delta \Pi_{WW}(M^2_W)-\Delta \Pi_{WW}(0)}{M^2_W},
\end{eqnarray}
where $\alpha(M^2_Z)$ is the fine-structure constant at the scale, $M_Z$, and $(s_W,c_W)=(\sin \theta_W,\cos \theta_W)$ are defined by the Weinberg angle, $\theta_W$. 
$S_{\rm 2HDM}$, $T_{\rm 2HDM}$ and $U_{\rm 2HDM}$ are the parameters in the ordinary 2HDMs,  which could be found in Refs.~\cite{EWPO-2HDM,EWPO-2HDM2}.
The new gauge boson $Z_H$ and the extra scalar boson $\tilde{h}$ in our model make new  
one-loop contributions to the vacuum polarizations of gauge fields, denoted by   
$(\Delta \Pi_{WW,ZZ})$.  
Their explicit expressions up to the $O(\xi)$ corrections are given by  
\begin{eqnarray}\label{eq:loop}
\Delta \Pi_{WW} (k^2) &=&\frac{\alpha}{4 \pi s^2_W}  
\{(c^{H^+_L}_{\phi_i} c^{H_l}_{h_i} 
c^{H^+_L}_{\phi_j} c^{H_l}_{h_j} )
B_{22} (k^2; m^2_{H_l},m^2_{H^+_L})
\nonumber\\
&&- \cos^2 (\beta-\alpha) B_{22} (k^2; m^2_{H},M^2_{W}) - \sin^2 (\beta-\alpha) B_{22} (k^2; m^2_{h},M^2_{W})   \nonumber \\
&&- \sin^2 (\beta-\alpha) B_{22} (k^2; m^2_{H},m^2_{H^+}) - \cos^2 (\beta-\alpha) B_{22} (k^2; m^2_{h},m^2_{H^+}) 
 \nonumber \\
&&+\frac{\gamma^2}{1+\gamma^2} B_{22} (k^2; m^2_{H^+},M^2_{Z_H})-\frac{\gamma^2}{1+\gamma^2} B_{22} (k^2; m^2_{H^+},m^2_{A}) \nonumber \\
&&
- M^2_W\left( \frac{v_i v_j}{v^2} c^{H_l}_{h_i} c^{H_l}_{h_j}\right) 
B_{0} (k^2; M^2_{W},m^2_{H_l})  
  \nonumber \\
&&+ M^2_W \cos^2 (\beta-\alpha)B_{0} (k^2; M^2_{W},m^2_{H})  + M^2_W \sin^2 (\beta-\alpha)B_{0} (k^2; M^2_{W},m^2_{h})  \}   \nonumber \\
&&   - M^2_W \frac{\alpha_H}{4 \pi} \left( \frac{v_i}{v}q_{H_i} c^{H^+_l}_{\phi_i} \right)^2   B_0(k^2; m_{H^+_l}^2,M_{Z_H}^2), \\
\Delta \Pi_{ZZ} (k^2) &=&\frac{\alpha}{4 \pi s^2_W c^2_W} 
  \{ (c^{A_m}_{\chi_i} c^{H_l}_{h_i} c^{A_m}_{\chi_j} c^{H_l}_{h_j}) 
B_{22} (k^2; m^2_{A_m},m^2_{H_l})
\nonumber \\
&&- \cos^2 (\beta-\alpha) B_{22} (k^2; M^2_{Z},m^2_{H}) - \sin^2 (\beta-\alpha) B_{22} (k^2; 
M^2_{Z},m^2_{h})   \nonumber \\
&&- \sin^2 (\beta-\alpha) B_{22} (k^2; m^2_{A},m^2_{H}) - \cos^2 (\beta-\alpha) B_{22} (k^2; 
m^2_{A},m^2_{h})  \nonumber \\
&&- M^2_Z 
\left  ( \frac{v_i v_j}{v^2} c^{H_l}_{h_i} c^{H_l}_{h_j}\right)
B_{0} (k^2; M^2_{Z},m^2_{H_l}) \nonumber \\
&&+ M^2_Z \cos^2 (\beta-\alpha) B_{0} (k^2; M^2_{Z},m^2_{H})
 + M^2_Z \sin^2 (\beta-\alpha) B_{0} (k^2; M^2_{Z},m^2_{h}) \} \nonumber \\
&& - M^2_Z \frac{\alpha_H}{ \pi} \left( \frac{v_i}{v}q_{H_i} c^{H_l}_{h_i} \right)^2   B_0(k^2; m_{H_l}^2,M_{Z_H}^2),
\label{eq:loop2}
 \end{eqnarray}
 which are used for phenomenological analyses of the EWPOs. 
We have defined a new parameter $\gamma$ for convenience: 
\beq
\gamma=\frac{v}{v_{\Phi}} \cos \beta \sin \beta .
\eeq
The extra corrections additionally depend on the mixing ($\alpha_{1,2}$) among 
the CP-even scalar bosons, the mass of the extra scalar boson ($m_{\widetilde{h}}$), 
and the $Z_H$ mass and its gauge coupling, $(M_{Z_H},g_H)$.  
The explicit expressions of the functions $B_0$ and $B_{22}$ can be found in 
Ref.~\cite{EWPO-2HDM}.

\subsection{Analysis in 2HDMs with $U(1)_H$ Gauge Symmetry}
\label{sec:STU with mixing}
Here, we discuss the bounds from EWPOs in the 2HDMs 
with $U(1)_H$ Higgs gauge symmetry.
For the numerical analysis, we use the following input parameters:
$M_Z=91.1875$ GeV,   $M_W=80.381$ GeV,  $\sin^2 \theta_W= 0.23116$,
$\alpha(M_Z)=1/(127.944)$, and $m_h=126$ GeV.
According to the recent LHC results, the bounds on $S$, $T$, and $U$
parameters are given by~\cite{Baak:2012kk,Baak:2013ppa}
\beq
S=0.03 \pm 0.10,~T=0.05 \pm 0.12, ~ U=0.03 \pm 0.10,
\eeq
with $m^{\rm ref}_h=126$ GeV and $m^{\rm ref}_t=173$ GeV.
The correlation coefficients are $+0.89_{ST}$, $-0.54_{SU}$, and $-0.83_{TU}$.\footnote{Fixing $U=0$, 
$S=0.05 \pm 0.09$ and $T=0.08 \pm 0.07$ with the correlation coefficient $+0.91$.
}

\begin{figure}[!t]
\begin{center}
{\epsfig{figure=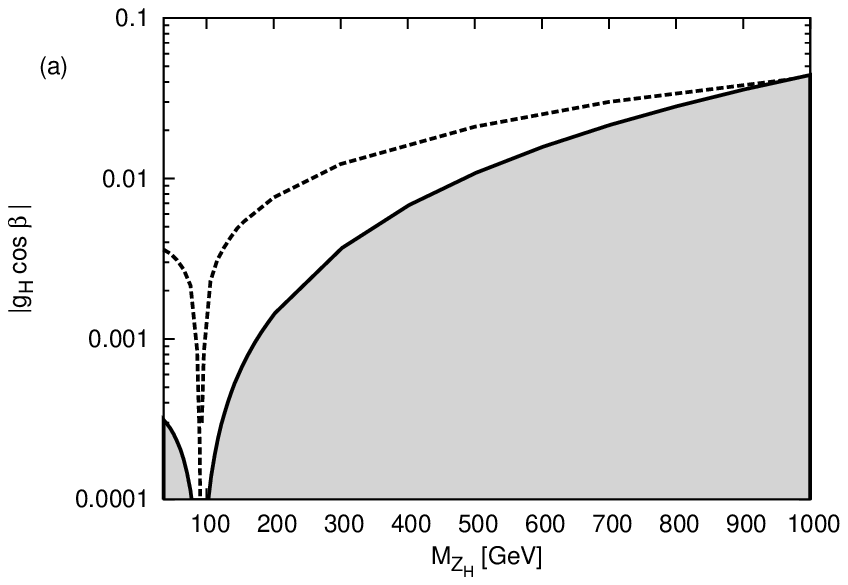,width=0.5\textwidth}}{\epsfig{figure=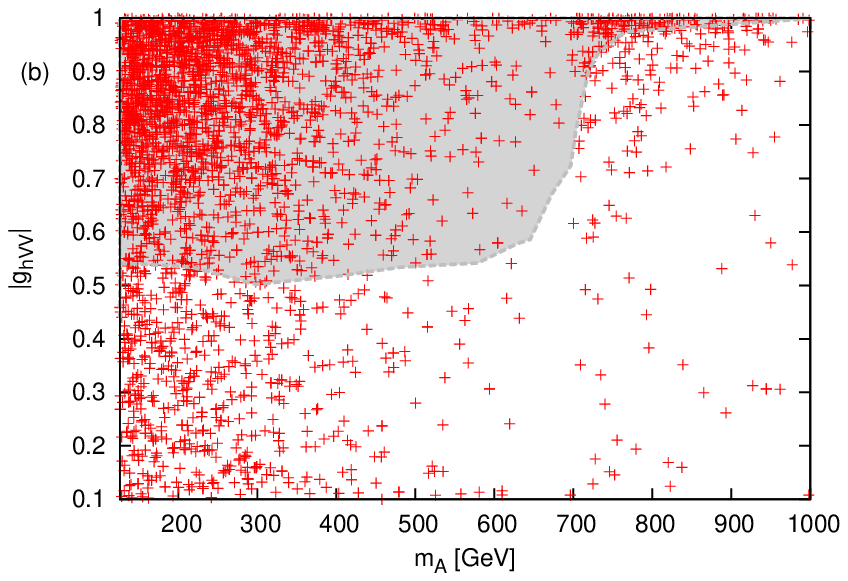,width=0.5\textwidth}}
\end{center}
\vspace{-0.5cm}
\caption{
Bounds on $M_{Z_H}$, $g_{H} \cos \beta $, $g_{hVV}$ and $m_A$ in the 2HDMs.
In the left panel, the gray region satisfies $\sin \xi \leq 10^{-3}$ coming from the collider experiments
while the dashed line is the upper limit coming from the $\rho$ parameter and $\Gamma_Z$. In the right panel,
the gray region is the allowed one for the type-I 2HDM with $g_{hVV}=\sin (\beta-\alpha)$ and $\alpha_1=\alpha_2=0$. 
The red points are allowed in the 2HDM with $U(1)_H$.
}
\label{fig:Bound-STU}
\end{figure}

\begin{figure}[!t]
\begin{center}
{\epsfig{figure=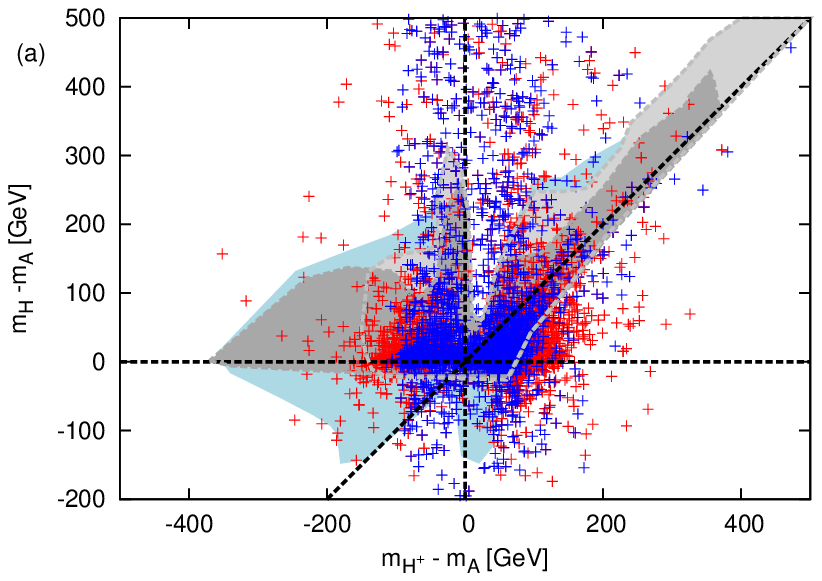,width=0.5\textwidth}}{\epsfig{figure=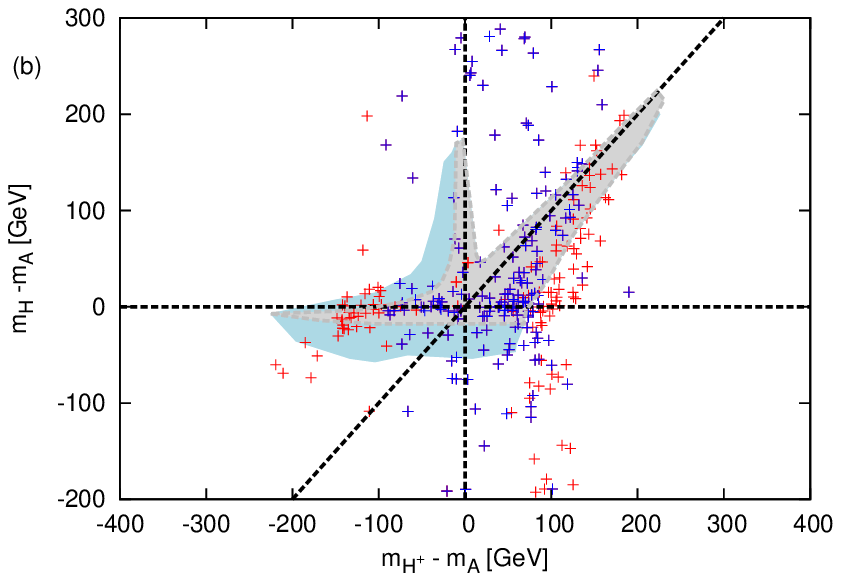,width=0.5\textwidth}}
\end{center}
\vspace{-0.5cm}
\caption{
Bounds on $m_{H^+}-m_A$ and $m_H-m_A$ in the 2HDMs. The gray (blue) regions are
allowed in the ordinary type-I 2HDM
(with $|\lambda_5| \leq 1$). In the left panel, $126$ GeV $\leq m_A$ $<700$ GeV is chosen and
the gray region is divided to the two mass regions: $126$ GeV $\leq m_A$ $<300$ GeV (gray) and 
$300$ GeV $\leq m_A$ $<700$ GeV (dark gray). In the right panel, $700$ GeV $\leq m_A$ $\leq 1000$ GeV is chosen.
The red (blue) points are allowed in the type-I 2HDM with $U(1)_H$ 
without (with) the conditions: $g_{hVV} \geq 0.9$ and $g_{htt} \geq 0.9$.
Three dashed lines corresponds to the ones for $m_{H^+}=m_A$, $m_{H}=m_A$, and $m_{H^+}=m_H$.
}
\label{fig:Bound-STU2}
\end{figure}

In Figs.~\ref{fig:Bound-STU} and \ref{fig:Bound-STU2}, the allowed regions within 
$90 \%$ C.L. of $S$, 
$T$, and $U$ parameters are presented in the type-I 2HDM with $q_{H_1}=q_{\Phi}=1$ 
and $q_{H_2}=0$.  
The parameters are scanned in the following regions: 
$1 \leq \tan \beta \leq 100$, $90$ GeV $\leq m_{H^+} \leq$ $1000$ GeV, 
$126$ GeV $\leq m_A, m_{H}, m_{\widetilde{h}} \leq$ $1000$ GeV, and $-1000$ GeV$ \leq \mu \leq$ $1000$ GeV. 
The constraints on the vacuum stability, unitarity and perturbativity 
introduced in the subsections~\ref{sec:vacuum stability} 
are imposed. The bound from $b \to s \gamma$ is assigned based on Ref.~\cite{Hermann:2012fc}.  
Light charged Higgs is constrained by the bound on exotic top decay $t \to H^+b$~\cite{topbound1,topbound2,topbound3} and 
the decay widths of $H, \widetilde{h} \to VV$ $(V=W,Z)$ are enough small 
to avoid the bounds 
in the collider experiments~\cite{Chatrchyan:2013yoa}.

%------------------------------------------------------------------------------
%------------------------------------------------------------------------------

In Figs.~\ref{fig:Bound-STU} (a) and (b), we show the bounds (a) on $M_{Z_H}$ and $g_{H} \cos \beta$ and 
(b) on $g_{hVV}=\sin (\beta-\alpha) \cos \alpha_1$ and $m_A$ in the type-I 2HDM with $U(1)_H$, respectively. 
Here $g_{hVV}$ is the $h$-$V$-$V$ $(V=W,Z)$ coupling normalized to the SM coupling.
In Fig.~\ref{fig:Bound-STU} (a), the gray region satisfies the collider bound, $\sin \xi \leq 10^{-3}$, 
mainly from the Drell-Yan process at the LHC and
the dashed line corresponds to the upper limit on the constrains coming from the $\rho$ parameter and 
$\Gamma_Z$.
In the region $M_{Z_H} \lesssim 1$ TeV, the collider bound is stronger than the bound from the $\rho$
parameter and $\Gamma_Z$.
We note that we include the one-loop corrections involving $Z_H$ to $S$, $T$, and $U$,
where $126$ GeV $\leq M_{Z_H} \leq$ $1000$ GeV and  $0$ $\leq |g_H| \leq$ $4 \pi$. 
The tree-level contribution to the $T$ parameter is also considered
but it just yields the deviation, $|\Delta T| \lesssim 0.01$.

In Fig.~\ref{fig:Bound-STU} (b), the gray region is allowed for 
$g_{hVV}$ and $m_A$ in the ordinary type-I 2HDM, where $\alpha_1=\alpha_2=0$ 
and $Z_H$ and $\Phi$ are decoupled.
If the pseudoscalar mass is heavy, 
$g_{hVV}$ should be close to one so that the Higgs signal around $126$ GeV should be SM-like.
The red points are allowed in the 2HDM with $U(1)_H$ with $\sin\xi\leq 10^{-3}$.
We note that the small $g_{hVV}$ region is also allowed 
due to an extra factor $\cos\alpha_1$ in $g_{hVV}$.
The small $g_{hVV}$ would reduce the production rate of the SM-like Higgs boson and 
the partial decay width of $h$ to the EW gauge bosons. 

%------------------------------------------------------------------------------
%------------------------------------------------------------------------------

In Fig.~\ref{fig:Bound-STU2}, we show the bounds on the mass differences among
$m_A$, $m_H$ and $m_{H^+}$. In Fig.~\ref{fig:Bound-STU2} (a), $m_A$ is less than $700$ GeV,
and the (dark) gray region satisfies $126$ GeV $\leq m_A$ $<300$ GeV ($300$ GeV $\leq m_A$ $<700$ GeV).
In Fig.~\ref{fig:Bound-STU2} (b), $m_A$ is within $700$ GeV $\leq m_A$ $\leq 1000$ GeV.
The gray region is allowed for all the constraints in the ordinary type-I 2HDM
with $\alpha_1=\alpha_2=0$ and $\lambda_5=0$. 
As we see in Appendix \ref{sec:Mass Matrix of CP-even scalars}, we can realize such small mixings assuming very small $\widetilde{\lambda_1}$,
$\widetilde{\lambda_2}$ and $\mu$ or very large $v_{\Phi}$.

The light blue region corresponds to the ordinary 2HDM with non-zero $\lambda_5$ ($|\lambda_5 | \leq 1$).
In the case of the 2HDMs with $\lambda_5=0$, the vacuum stability requires the relation (\ref{eq:relation-2HDM-0}).
On the other hand, non-zero $\lambda_5$ modifies the relation and, 
especially, negative $\lambda_5$ pushes the lower bound on $m_H$ down, so that the wider region is allowed 
in Figs.~\ref{fig:Bound-STU} and \ref{fig:Bound-STU2}. 

As we see in Fig.~\ref{fig:Bound-STU2}, each scalar mass could become different. 
However, it seems that at least two of them should be close to each other in the typical 2HDM 
with small $\lambda_5$. The heavier pseudoscalar mass requires the smaller mass difference.

In our 2HDM with $\widetilde{h}$ and $Z_H$,
the strict bounds could be evaded because of 
the contributions of the extra particles.
The red and blue points are allowed in the type-I 2HDM with $U(1)_H$ and
 the additional constraints, $g_{hVV}  \geq 0.9$ and $g_{htt}\geq 0.9$, are imposed on the blue points.
 Here $g_{htt}$ is the $h$-$t$-$\ov{t}$ coupling normalized to the SM coupling 
and it is given by $g_{htt}=\cos \alpha_1 \cos\alpha/\sin\beta$ in the type-I 2HDM with $U(1)_H$.
Once $\Phi$ is added and $h_{\Phi}$ mixes with $h_1$ and $h_2$,
the relation (\ref{eq:relation-2HDM-0}) is discarded, so that the red (blue) 
points exist outside of the gray region, when $h_{\Phi}$ and $Z_H$ reside in the $O(100)$ GeV scale. 
In particular, the predictions of the masses of the CP-even scalars 
are modified, so that
$m_H-m_A$ would have larger allowed region, compared with $m_{H^+}-m_A$.
Even if the SM Higgs search limits the normalized $h$-$V$-$V$ and $h$-$t$-$\ov{t}$ couplings,
the mass difference could not be constrained strongly as shown in the region of the blue points.

The constraints from EWPOs could easily be applied to the other type 2HDMs
by changing the experimental constraints on the charged Higgs mass. 
For example, $b \to s \gamma$
gives the lower bound on $m_{H^+} \gtrsim 360$ GeV in the type-II 2HDMs \cite{Hermann:2012fc}.

%-----------------------------------------------------------------------------------------------------------
\section{ Collider phenomenology of the Higgs bosons\label{sec:collider}}
%-----------------------------------------------------------------------------------------------------------
\subsection{Analysis strategies}
In this section, we consider collider phenomenology of the Higgs bosons,
in particular, focusing on the SM-like Higgs boson.
For the calculation of the decay rates of the neutral Higgs bosons,
we use the HDECAY~\cite{Djouadi:1997yw} 
with corrections to Higgs couplings to the SM fermions and gauge
bosons and with inclusion of the charged Higgs contribution to
the $h\to \gamma\gamma$ and $h\to Z\gamma$ decays.

There are 10 parameters in the potential neglecting the $Z_H$ boson effects 
at the EW scale, and one of them is fixed by the SM-like Higgs boson mass 
$m_h \sim 126$ GeV. We choose the other 9 parameters as 
$\tan \beta$, $m_A$, $d m_{H^+}$, $d m_H$, $m_{\tilde{h}}$,
$\alpha$, $\alpha_1$, $\alpha_2$, and $v_\phi$,
where $d m_{H^+}(d m_H ) = m_{H^+}(m_H) - m_A$ is the mass difference between
the charged Higgs (heavy Higgs) and pseudoscalar Higgs boson. 
In this analysis, we choose each parameter region as follows:
$1 \le \tan \beta \le 100$, 
126 GeV $\le m_A \le$ 1 TeV,
$| d m_{H^+, H} | \le 200$ GeV,
$ 0 \le \alpha, \alpha_{1,2} \le 2 \pi$,
126 GeV $\le m_{\tilde{h}} \le$ 1 TeV,
0 GeV $\le v_{\Phi} \le $ 3 TeV, respectively.\footnote{The larger mass-scale region 
could be considered, but they relate to the SM-Higgs signals indirectly 
through the bounds from the EWPOs and theoretical constraints, as we discuss in Sec. \ref{sec:EWPO}. 
Hence, they would not change our results in this section.}

In order to compare our models with the Higgs data at the LHC, we consider
the signal strength $\mu$ for each decay mode $i$ of the SM-like Higgs boson
with the production tag $j$, which is defined by
%-----------------
\begin{equation}
\mu^{i}_{j} = \frac{\sigma(pp\to h)^j_\textrm{2HDM} 
                    \textrm{Br}(h\to i)_\textrm{2HDM}}
                   {\sigma(pp\to h)^j_\textrm{SM} 
                    \textrm{Br}(h\to i)_\textrm{SM}},
\end{equation}
%-----------------
where $\sigma(pp\to h)^j$ means the production cross section for the SM-like Higgs boson
with the production tag $j$ and $\textrm{Br}(h\to i)$ is the branching ratio
of the SM-like Higgs boson decay into the $i$ state. Here $j=gg$, $Vh$, or $VVh$, which 
correspond to the $gg$ fusion production, vector boson associated production, and vector boson fusion production tag, respectively.  Finally $i = \gamma\gamma$, $WW$, $ZZ$,
or $\tau\tau$, depending on the decay channels.

The search for the SM Higgs boson also constrains the mass and couplings of the heavy Higgs boson.  In  high mass region greater than 200 GeV, the main search mode  is 
$h\to ZZ \to 4l$~\cite{heavyhiggs}.
For the SM-like Higgs boson, the lower limit for the Higgs boson mass is about 650 GeV 
and 300 GeV for the $gg$ fusion production and $VVh+Vh$ production, respectively. 
More detailed analysis is given  in Ref.~\cite{heavyhiggs}. 
For a Higgs boson $H$, the upper bound on the signal strength,
$\mu_{VVH,VH}^{ZZ}$ in the $VVH+VH$ production is about one or less for
$m_H > 200$ GeV while the bound on $\mu_{gg}^{ZZ}$ is about $0.1\sim 1$
for $200~\textrm{GeV} < m_H < 1$ TeV in the $gg$ fusion production, which
varies according to $m_H$. From the SM Higgs search for $m_H \le 200$ GeV,
we get the constraint on the signal strength
$\mu_{gg}^{ZZ} < 0.1 \sim 0.5$ whose bound
depends on $m_H$. We impose these bounds on the heavy Higgs boson $H$.

%For a SM-like Higgs boson, the upper bound on the signal strength
%in the $VVh+Vh$ production is about one or less for $m_H > 200$ GeV 
%while it is less than about $0.1$ for $200~\textrm{GeV} < m_H < 400$ GeV 
%in the $gg$ fusion production.  We impose this bound on the heavy Higgs boson $H$.

In this work, we consider two distinct cases in our Type-I 2HDM with $U(1)_H$ 
gauge symmetry:
\begin{itemize}
\item First, we consider the Type-I 2HDM with $U(1)_H$, 
assuming that $Z_H$ is decoupled from  the low energy Higgs physics. 
Then, the extra contribution is from only the extra Higgs scalar,
and the effect is parametrized by $m_{\widetilde{h}}$ and $\alpha_{1,2}$.
\item Secondly, we consider the Type-I 2HDM with $U(1)_H$, including $Z_H$
contribution. The charge assignment is fermiophobic by setting $u=d=0$.
In this case the $Z_H$ boson couples with the SM fermions only through the $Z$-$Z_H$ mixing,
and it contributes to the EWPOs.
\end{itemize}
We compare each case with the ordinary type-I 2HDM by setting $\alpha_{1,2}=0$ 
and omitting the singlet scalar $\Phi$ and $Z_H$.
We note  there is no $\lambda_5$ term in the Higgs potential in this case, 
as we mentioned in the  previous section.

%------------------------------------------------------------------------------
\begin{figure}[!t]
\begin{center}
{\epsfig{figure=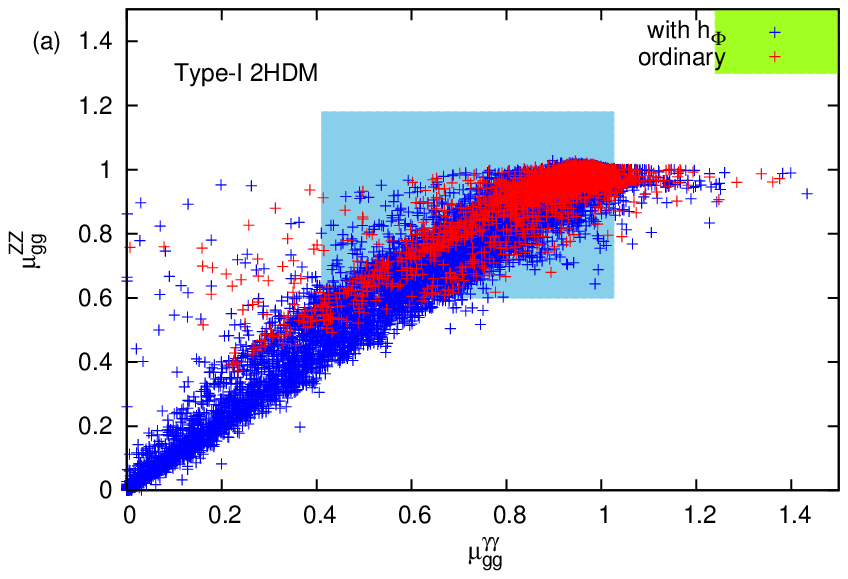,width=0.45\textwidth}}
{\epsfig{figure=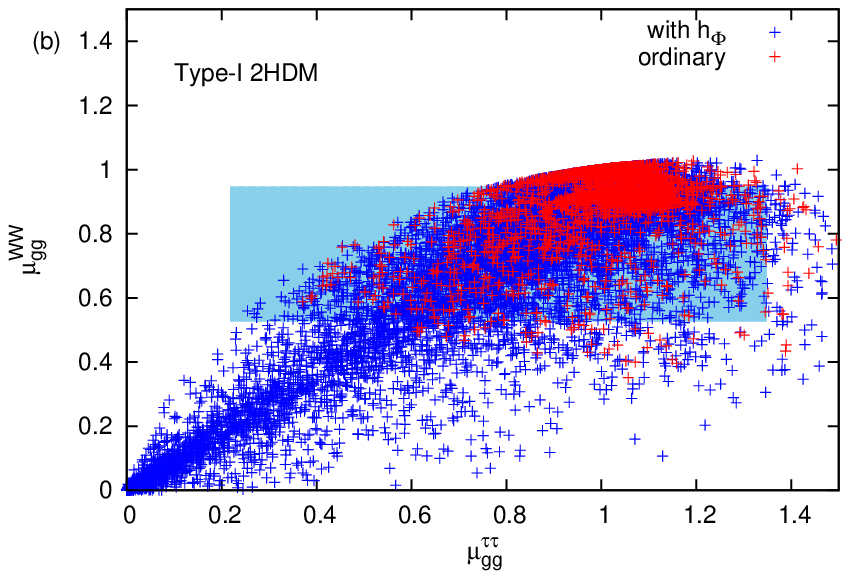,width=0.45\textwidth}}
\end{center}
\vspace{-0.5cm}
\caption{
(a) $\mu_{gg}^{\gamma\gamma}$ vs.~$\mu_{gg}^{ZZ}$ and
(b) $\mu_{gg}^{\tau\tau}$ vs.~$\mu_{gg}^{WW}$ in the ordinary type-I 2HDM (red)
and type-I 2HDM with $\Phi$ (blue).
The effect of $Z_H$ boson is assumed to be small enough to be ignored.
The skyblue and green regions are the allowed ones at CMS and ATLAS
in the 1$\sigma$ level.
}
\label{fig:gg2HDM}
\end{figure}
%------------------------------------------------------------------------------

\subsection{2HDM with the extra singlet scalar}
In this section, we consider the type-I 2HDM with the extra singlet scalar field, $h_\Phi$,
where we assume that the imaginary part of $\Phi$ is eaten by $Z_H$ and
 the effects of the $U(1)_H$ gauge boson are small enough to be ignored.
This could easily be achieved with an assumption of the heavy $Z_H$ mass 
and small $g_H$, namely in the limit of large $v_\Phi$.

We show the scattered plots for $\mu_{gg}^{\gamma\gamma}$ and $\mu_{gg}^{ZZ}$ 
in Fig.~\ref{fig:gg2HDM}(a), and for $\mu_{gg}^{\tau\tau}$ and $\mu_{gg}^{WW}$ in 
Fig.~\ref{fig:gg2HDM}(b), respectively.
The red points are allowed in the ordinary type-I 2HDM, whereas 
the blue points are consistent with the type-I 2HDM with $h_\Phi$, respectively.
The skyblue and green regions are consistent with the Higgs signal strengths reported 
by  CMS and ATLAS Collaborations within the 1$\sigma$ range, respectively, where
$\mu_{gg,\textrm{CMS}}^{\gamma\gamma}=0.70^{+0.33}_{-0.29}$,
$\mu_{gg,\textrm{ATLAS}}^{\gamma\gamma}=1.6\pm0.4$,
$\mu_{gg,\textrm{CMS}}^{ZZ}=0.86^{+0.32}_{-0.26}$,
and $\mu_{gg,\textrm{ATLAS}}^{ZZ}=1.8^{+0.8}_{-0.5}$.
Each signal strength at CMS is consistent with that at ATLAS within the $2\sigma$'s.

%The SM point is $\mu_{gg}^{\gamma\gamma,ZZ,WW,\tau\tau}=1$, which is in agreement 
%with the CMS data, but the ATLAS data are consistent only at the $2\sigma$ level.
%In the ordinary 2HDM, the allowed points are in the regions of  
%$0.8 \lesssim \mu_{gg}^{\gamma\gamma} \lesssim 1.2$ and 
%$0.6 \lesssim \mu_{gg}^{ZZ} \lesssim 1.1$.
%However, in the 2HDM with $h_\Phi$ the allowed region is much wider:   
%$0 \lesssim \mu_{gg}^{\gamma\gamma} \lesssim 1.2$ and 
The SM point is $\mu_{gg}^{\gamma\gamma,ZZ,WW,\tau\tau}=1$, which is in agreement 
with the CMS data, but the ATLAS data are consistent only at the $2\sigma$ level. 
In the ordinary 2HDM, the allowed points are in the regions of  
$\mu_{gg}^{\gamma\gamma} \lesssim 1.4$ and 
$0.4 \lesssim \mu_{gg}^{ZZ} \lesssim 1.1$.
In the 2HDM with $h_\Phi$ the allowed region is wider
in the $gg\to h\to ZZ$ process:   
$0 \lesssim \mu_{gg}^{ZZ} \lesssim 1.1$.
Both 2HDMs contain the SM point $\mu = 1$, and the CMS data for 
$\mu_{gg}^{\gamma\gamma}$ and $\mu_{gg}^{ZZ}$, but  only the edge of the allowed 
region is barely consistent with the ATLAS data in the $2\sigma$ level. 

%For $\mu_{gg}^{\tau\tau}$ both models predict a large allowed region from 0 ($0.3$) to 
For $\mu_{gg}^{\tau\tau}$ both models predict a large allowed region from 0 ($0.4$) to 
$1.5$ or larger so that it is difficult to constrain the parameters in the 2HDMs using only 
$\mu_{gg}^{\tau\tau}$.

%In the ordinary 2HDM $0.6 \lesssim \mu_{gg}^{WW} \lesssim 1$ is allowed,  whereas much 
%wider region  $0 \lesssim \mu_{gg}^{WW} \lesssim 1$ is allowed in the 2HDM with 
%$h_\Phi$.   The allowed region in the 2HDM with $h_\Phi$ is much broader than
%that in the ordinary 2HDM.  

%As shown in Fig~\ref{fig:gg2HDM} (a), the region of
%$\mu_{gg}^{\gamma\gamma} \lesssim 0.8$ and $\mu_{gg}^{ZZ} \lesssim 0.6$
%is not allowed in the ordinary 2HDM.
%Hence, if it turns out that the two signal strengths are less than $0.6$,
%one might conclude that the 2HDM with $h_\Phi$ is more favored
%than the ordinary 2HDM.
%However, if it turns out that each signal strength is close
%to the SM point, the 2HDM with
%$h_\Phi$ cannot be distinguished from the ordinary 2HDM  as well as the SM.
%The mixing with the extra CP-even scalar decreases the two signal strengths, 
%so that we could conclude that their upper bounds are
%$\mu_{gg}^{\gamma\gamma} \lesssim 1.2$ and $\mu_{gg}^{ZZ} \lesssim 1.0$ 
%in the type-I 2HDM with the extra scalar.

In the ordinary 2HDM $0.4 \lesssim \mu_{gg}^{WW} \lesssim 1$ is allowed,  whereas much 
wider region  $0 \lesssim \mu_{gg}^{WW} \lesssim 1$ is allowed in the 2HDM with $h_\Phi$.  
The allowed region in the 2HDM with $h_\Phi$ is much broader than that in the ordinary 2HDM.  

As shown in Fig~\ref{fig:gg2HDM}, the region of
$\mu_{gg}^{ZZ} \lesssim 0.4$ and $\mu_{gg}^{WW} \lesssim 0.4$
is not allowed in the ordinary 2HDM.
Hence, if it turns out that the two signal strengths were less than $0.4$,
one might be able to conclude that the 2HDM with $h_\Phi$ is more favored
than the ordinary 2HDM.  However, if it turns out that each signal strength is close
to the SM point, the 2HDM with $h_\Phi$ cannot be distinguished from the ordinary 2HDM  
as well as the SM. 
The mixing with the extra CP-even singlet scalar decreases the two signal strengths, 
so that we could conclude that their upper bounds are
$\mu_{gg}^{\gamma\gamma} \lesssim 1.4$ and $\mu_{gg}^{ZZ} \lesssim 1.0$ 
in the type-I 2HDM with the extra scalar.

%------------------------------------------------------------------------------
\begin{figure}[!t]
\begin{center}
{\epsfig{figure=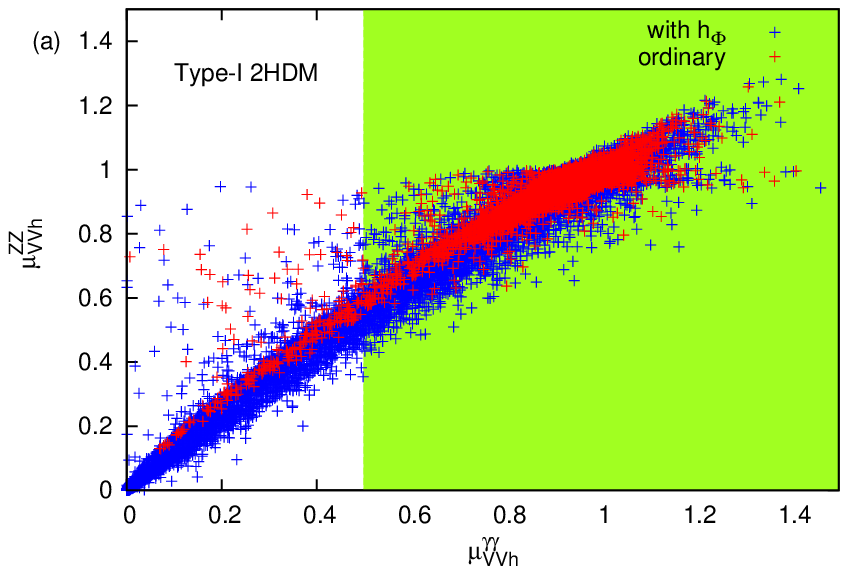,width=0.45\textwidth}}
{\epsfig{figure=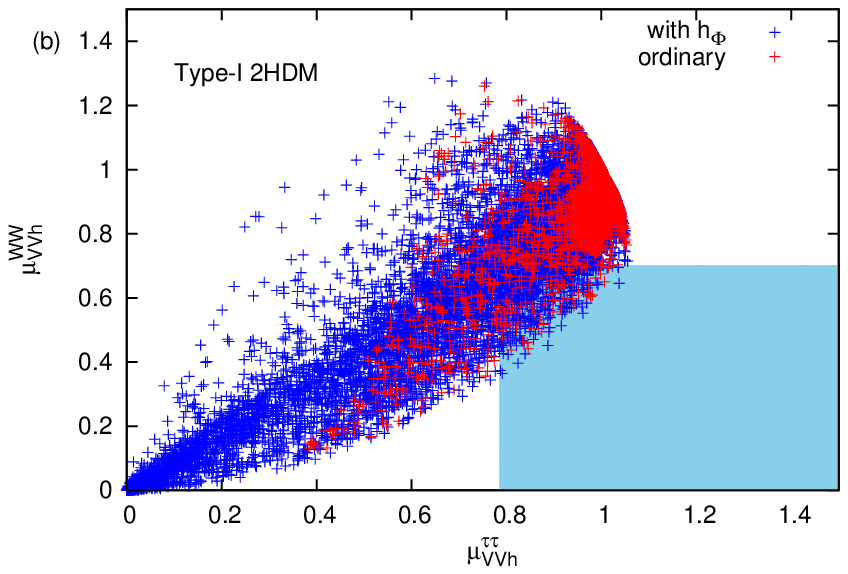,width=0.45\textwidth}}
{\epsfig{figure=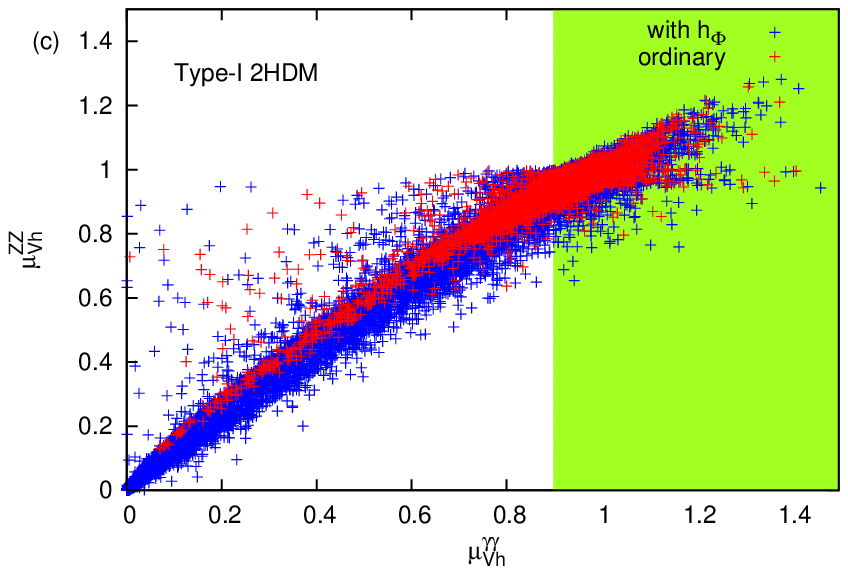,width=0.45\textwidth}}
{\epsfig{figure=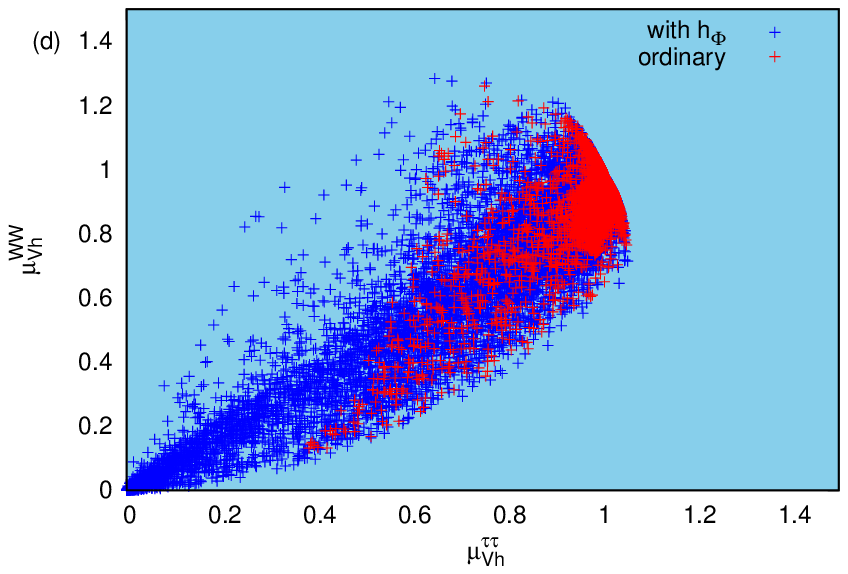,width=0.45\textwidth}}
\end{center}
\vspace{-0.5cm}
\caption{
(a) $\mu_{VVh}^{\gamma\gamma}$ vs.~$\mu_{VVh}^{ZZ}$,
(b) $\mu_{VVh}^{\tau\tau}$ vs.~$\mu_{VVh}^{WW}$ 
(c) $\mu_{Vh}^{\gamma\gamma}$ vs.~$\mu_{Vh}^{ZZ}$, and
(d) $\mu_{Vh}^{\tau\tau}$ vs.~$\mu_{Vh}^{WW}$ 
in the ordinary type-I 2HDM (red)
and type-I 2HDM with $h_\Phi$ (blue).
The effect of $Z_H$ boson is assumed to be small enough to be ignored.
The skyblue and green regions are the allowed ones at CMS and ATLAS
in the 1$\sigma$ level.
}
\label{fig:vv2HDM}
\end{figure}
%------------------------------------------------------------------------------

Fig.~\ref{fig:vv2HDM} shows the scattered plots 
(a) for $\mu_{VVh}^{\gamma\gamma}$ and $\mu_{VVh}^{ZZ}$,
(b) for $\mu_{VVh}^{\tau\tau}$ and $\mu_{VVh}^{WW}$,
(c) for $\mu_{Vh}^{\gamma\gamma}$ and $\mu_{Vh}^{ZZ}$, and
(d) for $\mu_{Vh}^{\tau\tau}$ and $\mu_{Vh}^{WW}$, respectively.
The red points are allowed in the ordinary type-I 2HDM, while  the blue ones are in the 
type-I 2HDM with $h_\Phi$.
In the SM, $\mu_{VVh,Vh}^{\gamma\gamma,ZZ,WW,\tau\tau}=1$ is satisfied. 
In these figures, the experimental data are consistent with the SM prediction
at the $1\sigma$ level except $\mu_{VVh}^{WW}$. 
However, it does not imply any conclusive deviation from the SM 
since the experimental uncertainties are very large at the moment.
As shown in the figures, $\mu_{VVh,Vh}^{ZZ,WW}$ could get much larger than 
the SM prediction in the parameter regions which increase the branching ratios
of $h\to ZZ$ or $h\to WW$.
We note that the decay widths of the Higgs boson $h$ into $ZZ$ or $WW$ are
rescaled by $g_{hVV}=\cos \alpha_1 \sin(\beta-\alpha)$, while
those into a fermion pair are by $g_{hff}=\cos\alpha_1 \cos\alpha /\sin\beta$.
In the limit of small $\cos \alpha$ or large $\sin \beta$, the branching ratio of the 
$h$ decay into a $b\bar{b}$ pair could get much smaller than the branching ratio 
in the SM and as a result,  the branching ratios of the $h$ decay into $ZZ$ or $WW$ 
could be much enhanced.

%As shown in Fig~\ref{fig:vv2HDM}, the regions of
%$\mu_{VVh,Vh}^{\gamma\gamma} \lesssim 0.6$, $\mu_{VVh,Vh}^{ZZ,WW} \lesssim 0.6$
%and $\mu_{VVh,Vh}^{\tau\tau} \lesssim 0.6$
%are not allowed in the ordinary 2HDM.
%Hence, if it turns out that the signal strengths are less than $0.6$,
%one might conclude that the 2HDM with $h_\Phi$ is more favored
%than the ordinary 2HDM.
%In the region of $\mu>0.6$ we cannot distinguish the 2HDM with $h_\Phi$ 
%from the ordinary 2HDM. 
As shown in Fig~\ref{fig:vv2HDM}, the region of
$\mu_{VVh,Vh}^{\tau\tau} \lesssim 0.4$
is not allowed in the ordinary 2HDM.
Hence, if it turns out that the signal strengths are less than $0.4$,
one might conclude that the 2HDM with $h_\Phi$ is more favored
than the ordinary 2HDM.
In the region of $\mu_{VVh,Vh}>0.4$ we cannot distinguish the 2HDM with $h_\Phi$ 
from the ordinary 2HDM. 
If it turns out that each signal strength is close
to the SM point, the 2HDM with $h_\Phi$ 
cannot be distinguished from the ordinary 2HDM  as well as the SM.

%------------------------------------------------------------------------------
\begin{figure}[!t]
\begin{center}
{\epsfig{figure=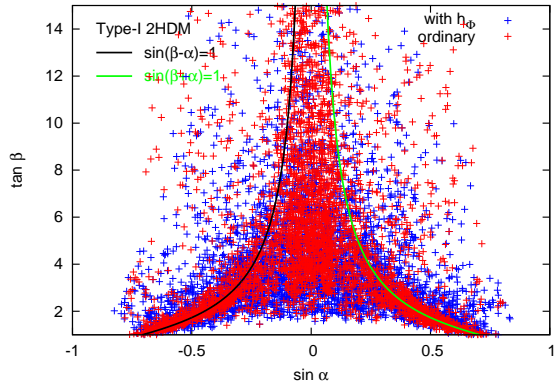,width=0.5\textwidth}}
\end{center}
\vspace{-0.5cm}
\caption{
$\sin \alpha$ vs. $\tan \beta$ in the type-I ordinary 2HDM (red) and 
in the type-I 2HDM with $h_\Phi$ (blue). The points are consistent with
the CMS data for $\mu_{gg}^{\gamma\gamma}$ and $\mu_{gg}^{ZZ}$
in the $1\sigma$ level.
The black and green lines correspond to the cases 
$\sin(\beta-\alpha)=1$ (SM limit) and $\sin(\beta+\alpha)=1$, respectively.
}
\label{fig:angles2HDM}
\end{figure}
%------------------------------------------------------------------------------

In Fig.~\ref{fig:angles2HDM}, we depict the scattered plot 
for $\sin \alpha$ and $\tan \beta$. The red and blue points are consistent
with the CMS data for $\mu_{gg}^{\gamma\gamma}$ and $\mu_{gg}^{ZZ}$
at the $1\sigma$ level in the type-I ordinary 2HDM and
in the type-I 2HDM with $h_\Phi$, respectively.
The black line corresponds to the SM limit $\sin(\beta-\alpha)=1$
while the green line to $\sin(\beta+\alpha)=1$.
%In the ordinary 2HDM, the allowed points are concentrated near the two lines, 
%$\sin(\beta \mp \alpha)=1$.
%However, in the 2HDM with $h_\Phi$ more parameter regions, in particular
%$\alpha \sim 0$ region, are allowed.
In the ordinary 2HDM and the 2HDM with $h_\Phi$,  
the allowed points are scattered over the region $|\sin\alpha| \lesssim 0.8$.
The region $|\sin\alpha| \gtrsim 0.8$ is forbidden, since the 
coupling $g_{hff} \sim \cos\alpha/\sin\beta$ to the 
fermions becomes small for $\tan\beta >1$.
In both models, the allowed regions contain the SM limit 
$\sin(\beta - \alpha)=1$ and there is no distinction between the two models.
There is no region which agrees with the ATLAS data 
for $\mu_{gg}^{\gamma\gamma}$ and $\mu_{gg}^{ZZ}$ at the $1\sigma$ level,
but one can obtain a similar figure for the ATLAS data at the $2\sigma$ level.

%------------------------------------------------------------------------------
\begin{figure}[!t]
\begin{center}
{\epsfig{figure=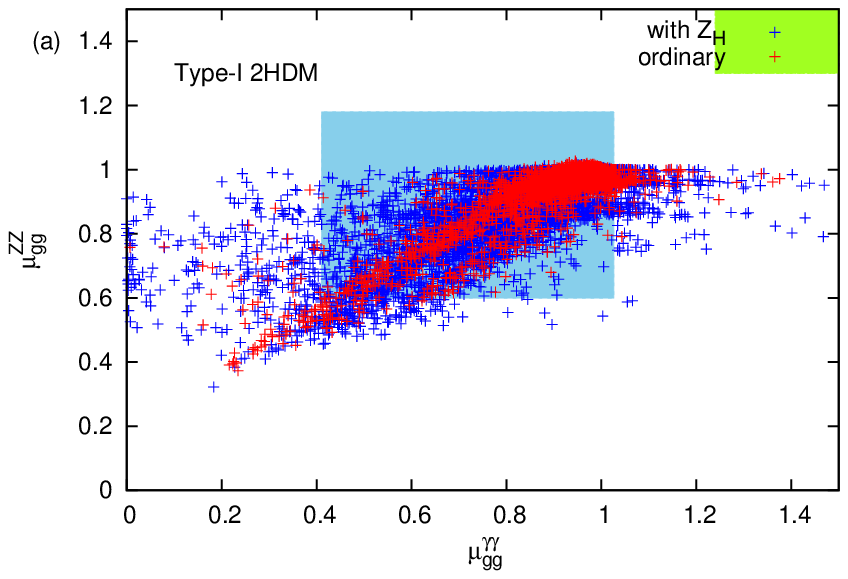,width=0.45\textwidth}}
{\epsfig{figure=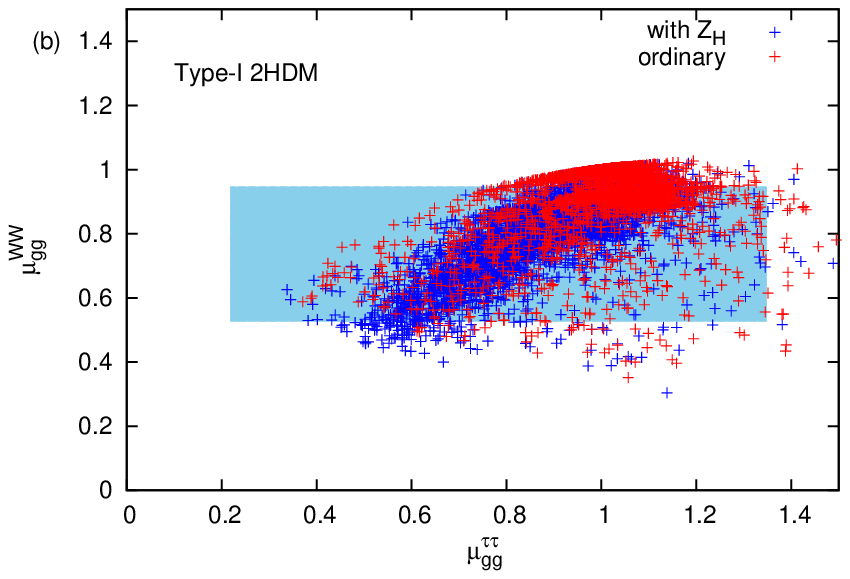,width=0.45\textwidth}}
\end{center}
\vspace{-0.5cm}
\caption{
(a) $\mu_{gg}^{\gamma\gamma}$ vs.~$\mu_{gg}^{ZZ}$ and
(b) $\mu_{gg}^{\tau\tau}$ vs.~$\mu_{gg}^{WW}$ in the ordinary type-I 2HDM (red)
and type-I 2HDM with a $Z_H$ (blue).
The skyblue and green regions are the allowed ones at CMS and ATLAS
in the 1$\sigma$ level.
}
\label{fig:gg2HDMzh}
\end{figure}
%------------------------------------------------------------------------------

%------------------------------------------------------------------------------
\begin{figure}[!t]
\begin{center}
{\epsfig{figure=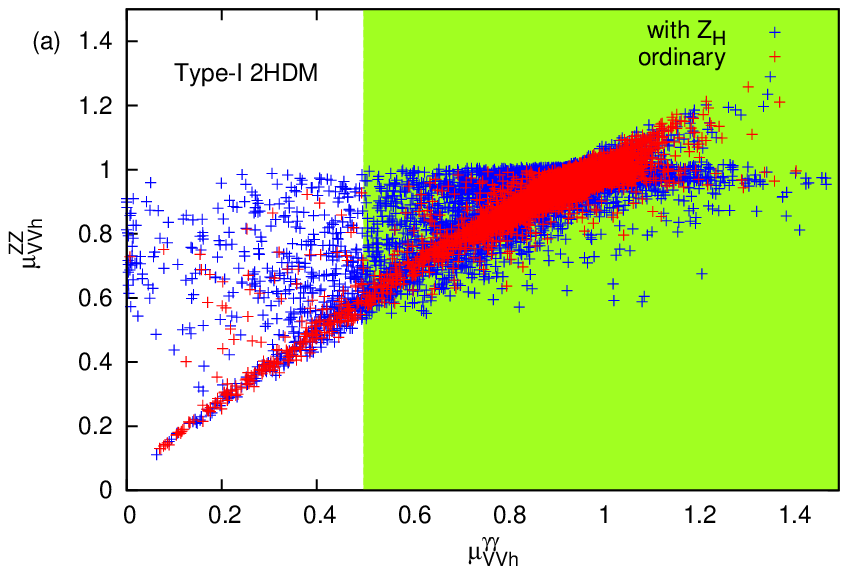,width=0.45\textwidth}}
{\epsfig{figure=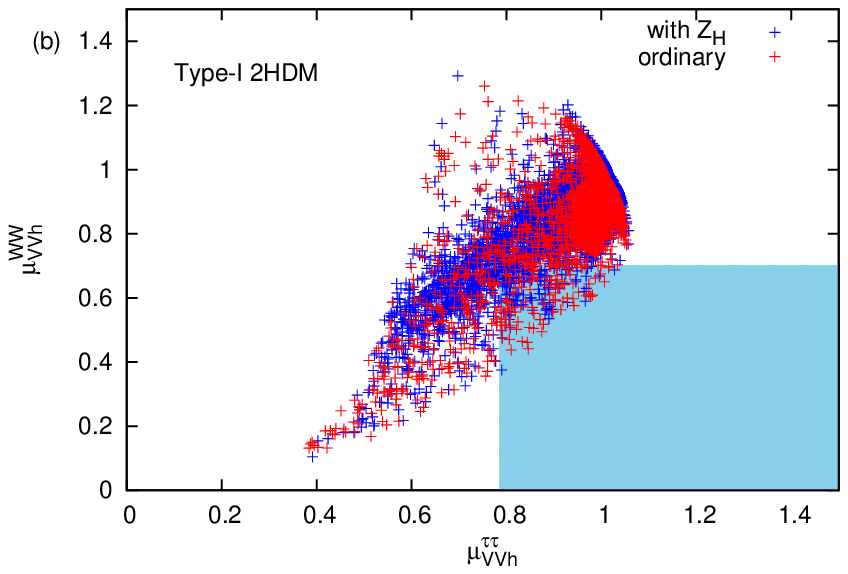,width=0.45\textwidth}}
{\epsfig{figure=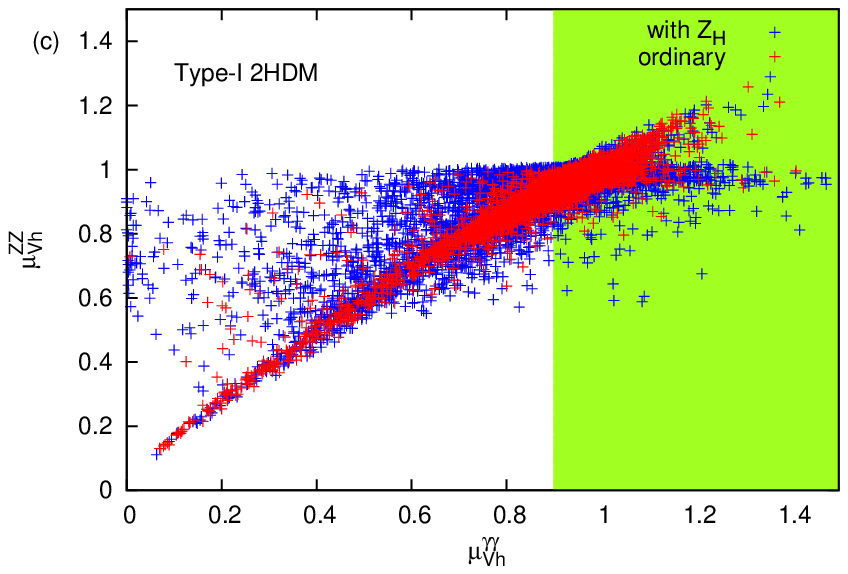,width=0.45\textwidth}}
{\epsfig{figure=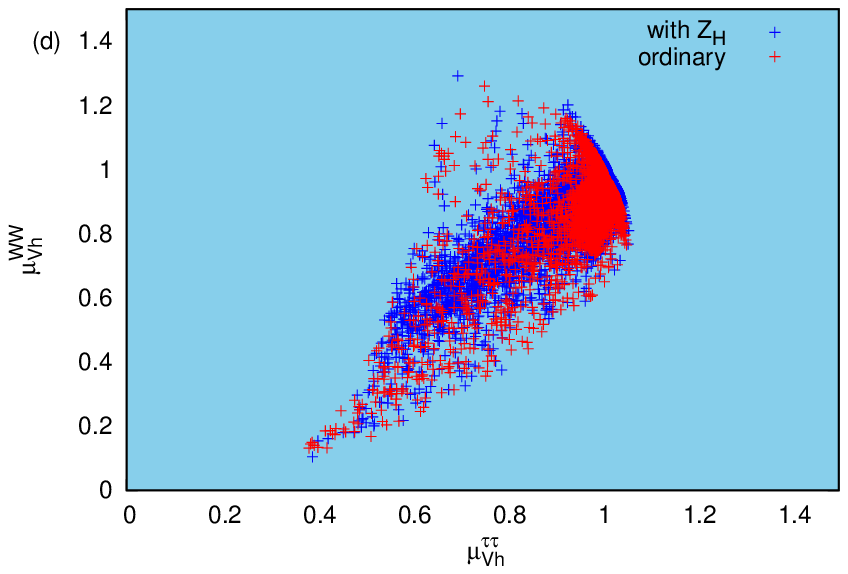,width=0.45\textwidth}}
\end{center}
\vspace{-0.5cm}
\caption{
(a) $\mu_{VVh}^{\gamma\gamma}$ vs.~$\mu_{VVh}^{ZZ}$,
(b) $\mu_{VVh}^{\tau\tau}$ vs.~$\mu_{VVh}^{WW}$ 
(c) $\mu_{Vh}^{\gamma\gamma}$ vs.~$\mu_{Vh}^{ZZ}$, and
(d) $\mu_{Vh}^{\tau\tau}$ vs.~$\mu_{Vh}^{WW}$ 
in the ordinary type-I 2HDM (red)
and type-I 2HDM with a $Z_H$ (blue).
The skyblue and green regions are the allowed ones at CMS and ATLAS
in the 1$\sigma$ level.
}
\label{fig:vv2HDMzh}
\end{figure}
%------------------------------------------------------------------------------

%------------------------------------------------------------------------------
\begin{figure}[!t]
\begin{center}
{\epsfig{figure=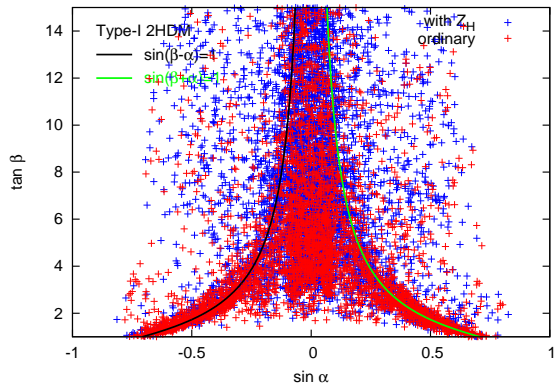,width=0.5\textwidth}}
\end{center}
\vspace{-0.5cm}
\caption{
$\sin \alpha$ vs. $\tan \beta$ in the type-I ordinary 2HDM (red) and 
in the type-I 2HDM with a $Z_H$ boson (blue). 
The points are consistent with
the CMS data for $\mu_{gg}^{\gamma\gamma}$ and $\mu_{gg}^{ZZ}$
in the $1\sigma$ level.
The black and green lines correspond to the cases 
$\sin(\beta-\alpha)=1$ (SM limit) and $\sin(\beta+\alpha)=1$, respectively.
}
\label{fig:angles2HDMzh}
\end{figure}
%------------------------------------------------------------------------------

\subsection{2HDM with the  $Z_H$ boson: fermiophobic case}
In this section, we discuss the 2HDM with $U(1)_H$ where the $U(1)_H$ gauge 
boson $\hat{Z}_H$ is fermiophobic, assuming $u=d=0$ as shown in Table~\ref{table1}.
Then the $\hat{Z}_H$ boson does not couple with the SM fermions, but
in the mass eigenstate the $Z_H$ boson, which is a mixture of $\hat{Z}$ and 
$\hat{Z}_H$, can couple with the SM fermions, and the couplings of the $Z$ boson 
is modified by the mixing angle between $\hat{Z}$ and $\hat{Z}_H$.

In this model, we have 10 parameters except $m_h$ fixed to 126 GeV.
The general model allows the mixing of $h_\Phi$, $h_1$ and $h_2$
as shown in Eq. (\ref{mixinghp}). However, the analysis of the model
is time-consuming and the general feature of mixing between two Higgs doublets 
and singlet fields would reduce signal strengths as in the previous section.
Therefore, we consider no mixing case by setting $\alpha_1=\alpha_2=0$
and compare our results with the typical 2HDM.

We choose the parameter regions as follows:
$1 \le \tan \beta \le 100$,
126 GeV $\le m_A \le$ 1 TeV,
$| d m_{H^+, H} | \le 200$ GeV,
$ 0 \le \alpha \le 2 \pi$,
126 GeV $\le m_{\widetilde{h}} \le$ 1 TeV.
The $U(1)_H$ coupling $g_H$ and the mass of $Z_H$ are chosen to be
$0 \le g_H \le \sqrt{4 \pi}$ and
$36~\textrm{GeV} \le M_{Z_H} \le 1$ TeV,
where the low bound for $M_{Z_H}$ is taken to suppress the decay mode
$h\to Z Z_H$. Then, $v_\Phi$ is given in terms of parameters:
$v_\Phi = [{M_{Z_H}^2/g_H^2 - v^2 / (1+\tan^2\beta)}]^{1/2}$.
In the range of $M_{Z_H} \le m_h/2$, $h$ can decay into $Z_H Z_H$.
However, in our $U(1)_H$ charge assignment $(1,1,0)$
on the Higgs fields $(\Phi, H_1, H_2)$, the branching ratio for $h\to Z_H Z_H$ is
suppressed. Actually for the parameters which pass all experimental constraints,
we find that $\textrm{Br}(h\to Z_H Z_H) < 10^{-5}$ which can be 
safely ignored in phenomenological analysis.

We depict the scattered plots for $\mu_{gg}^{\gamma\gamma}$ and 
$\mu_{gg}^{ZZ}$ in Fig.~\ref{fig:gg2HDMzh}(a), and for $\mu_{gg}^{\tau\tau}$
and $\mu_{gg}^{WW}$ in Fig.~\ref{fig:gg2HDMzh}(b), respectively.
The red and blue points correspond to the ordinary Type-I 2HDM
and Type-I 2HDM with the $Z_H$ boson, respectively.
The skyblue and green regions are CMS and ATLAS bounds at the $1\sigma$ level.
%As shown in Fig.~\ref{fig:gg2HDMzh}(a),
%in the ordinary 2HDM, $0.8 \lesssim \mu_{gg}^{\gamma\gamma} \lesssim 1.2$
%and $0.6 \lesssim \mu_{gg}^{ZZ} \lesssim 1.05$ are predicted,   
%but  in the 2HDM with the $Z_H$ boson, broader regions of the Higgs signal strengths 
%are allowed and $\mu_{gg}^{\gamma\gamma}$ and $\mu_{gg}^{ZZ}$ could be much 
%smaller than the SM prediction. 
As shown in Fig.~\ref{fig:gg2HDMzh},
the 2HDM with the $Z_H$ boson seems to have broader regions of the Higgs signal 
strengths than those in the ordinary 2HDM, but there is no essential difference.
In case of the general mixing between the neutral Higgs bosons, we might be able 
to distinguish the 2HDM with the $Z_H$ boson from the ordinary 2HDM in some 
parameter   spaces, especially  in the region $\mu_{gg}^{ZZ,WW} \lesssim 0.4$. 
However, this region is inconsistent with the current measurements.
Both 2HDMs are consistent with the CMS data at the $1\sigma$ level.
However, it is difficult to increase 
$\mu_{gg}^{\gamma\gamma}$ and $\mu_{gg}^{ZZ}$ to the ATLAS data in the present 
models.  Therefore the 2HDM with the $Z_H$ boson are not in agreement with the 
ATLAS data at the $1\sigma$ level.

Fig.~\ref{fig:vv2HDMzh} shows the scattered plots 
(a) for $\mu_{VVh}^{\gamma\gamma}$ and $\mu_{VVh}^{ZZ}$,
(b) for $\mu_{VVh}^{\tau\tau}$ and $\mu_{VVh}^{WW}$ 
(c) for $\mu_{Vh}^{\gamma\gamma}$ and $\mu_{Vh}^{ZZ}$, and
(d) for $\mu_{Vh}^{\tau\tau}$ and $\mu_{Vh}^{WW}$, respectively.
The red points are allowed in the ordinary type-I 2HDM while 
the blue ones are in the 2HDM with the $Z_H$ boson.
In the 2HDM with the $Z_H$ boson, 
$\mu_{VVh,Vh}^{ZZ,WW}$ could get much larger than 
the SM prediction as shown in the figures.
If the mixing between the two Higgs doublet and singlet fields are allowed,
broader region  with smaller  signal strengths  would be allowed as in the 2HDM with 
$h_{\Phi}$ discussed in the previous subsection.
The SM points $\mu_{VVh,Vh}^{ZZ,WW}=1$ are consistent with the (ordinary) 2HDMs
at the $1\sigma$ level except for $\mu_{VVh}^{WW}$.  
However  the deviation in $\mu_{VVh}^{WW}$ is not statistically significant yet 
because of large experimental errors.

In Fig.~\ref{fig:angles2HDMzh}, we depict the scattered plot 
for $\sin \alpha$ and $\tan \beta$, where the red and blue points are 
consistent with the CMS data for $\mu_{gg}^{\gamma\gamma}$ and $\mu_{gg}^{ZZ}$
in the $1\sigma$ level in the type-I ordinary 2HDM and
in the type-I 2HDM with the $Z_H$ boson, respectively.
The black line corresponds to the SM limit $\sin(\beta-\alpha)=1$
while the green line to $\sin(\beta+\alpha)=1$.
%In the 2HDM with the $Z_H$ boson, larger parameter regions, in particular
%$\alpha \sim 0$ region, are allowed, whereas in the ordinary 2HDM 
%only the regions near $\sin(\beta\pm\alpha)=1$ are allowed.
As in the 2HDM with $h_\Phi$, the region $|\sin\alpha|\gtrsim 0.8$ is
not allowed and there is no difference between the ordinary 2HDM
and the 2HDM with $U(1)_H$ Higgs gauge symmetry in the type-I case
even though the extra $Z_H$ boson contribution is taken into account.
However, in the type-II 2HDMs, one could find apparent distinction
between the 2HDMs without $U(1)_H$ Higgs gauge symmetry
and with the gauge symmetry~\cite{progress}.

%%%%%%%%%%%%%%%%%%%%%%%%%%%%%%%%%%%%%%%%
\section{Conclusion\label{sec:con}}

Discovery of a SM-like Higgs boson at the LHC has opened a new era 
in particle physics.  
It is imperative to answer the question if this new boson is the SM Higgs boson or one of 
Higgs bosons in an extended model with multi-Higgs  fields. 
The 2HDM is one of the simplest models which extend the SM Higgs sector and is well 
motivated by MSSM, GUT, etc.
In Ref.~\cite{Ko-2HDM}, it was suggested to replace the $Z_2$ symmetry in the ordinary 2HDM 
with $U(1)_H$ gauge symmetry, which can easily realize the NFC criterion 
with proper $U(1)_H$ charge assignments to the two Higgs doublets and the SM chiral fermions.
The local $U(1)_H$ symmetry may be the origin of softly broken $Z_2$ symmetry  
which has been widely discussed so far.

In this paper, we performed detailed phenomenological analysis of the observed 126 GeV Higgs boson within the Type-I 2HDM with the $U(1)_H$ symmetry proposed in Ref.~\cite{Ko-2HDM}.  
We added an extra complex scalar that breaks $U(1)_H$ spontaneously, 
in order to avoid the strong constraint on the mixing between the $Z$ boson and 
the extra $Z_H$ boson from EWPOs.
%{\color{blue} 
%Once the Higgs $SU(2)_L$ doublet is charged under gauge symmetry, the extra gauge 
%symmetry is also spontaneously broken at the same time as the EW symmetry breaking, 
%and then the mass mixing between the extra gauge boson and $Z$ boson appears in general. 
%It is strongly constrained by EWPOs, so that we added one extra complex scalar which 
%contributes to the $U(1)_H$ gauge symmetry breaking and  studied the strong bounds including 
%the theoretical bounds such as the vacuum stability. (CAN WE REMOVE THIS PART ?)}
Our extension of 2HDMs predicts one extra gauge boson and one extra neutral scalar compared with the 2HDMs with $Z_2$ symmetry, and allows a large pseudoscalar mass according to the spontaneous $U(1)_H$ symmetry breaking. 
%{\color{blue}
%If the mixing angles, $\alpha_{1,2}$, are negligibly small,  the allowed region becomes smaller 
%than in 2HDMs with softy broken $Z_2$ symmetry because $\lambda_5$ is absent.
%When the pseudoscalar is heavier than $700$ GeV, one mass difference among 
%$m_A$, $m_H$ and $m_{H^+}$ should be less than $\sim 200$ GeV, and the other should be 
%at most $O(10)$ GeV.   When the mixing is not negligible, especially the allowed region for 
%$m_H$ becomes wider even if the SM-Higgs signal  is SM-like.
% (CAN WE REMOVE THIS PART OR REDUCE ?)}

Taking into account experimental constraints from the SM-Higgs search, EWPO etc., 
and theoretical constraints from perturbativity, unitarity, and vacuum stability,
we studied the signal strengths in two different cases:
\begin{itemize}
\item  Case I: 
Type-I 2HDM with the extra scalar $h_\Phi$, assuming the $U(1)_H$ gauge boson is heavy 
enough to be decoupled at the EW scale. 
In this case,  the Higgs sector includes an extra scalar which 
is a remnant from spontaneous $U(1)_H$ symmetry breaking, and the EWPOs will be affected.
We found that the signal strengths in the 2HDMs with $h_\Phi$ could be much smaller 
than those in the 2HDM with $Z_2$ symmetry in some channels. 
However, if the signal strengths are close to the SM prediction,
it would be nontrivial to distinguish the 2HDM with $h_\Phi$ from the 2HDM with $Z_2$ 
symmetry with Higgs signal strengths alone, especially when all the signal strengths are 
observed close to the SM values.  
In case the signal strengths are bigger than the SM prediction,
the extra mixing of CP-even scalars does not help to save type-I 2HDM especially in 
$h \to VV$.
\item Case II: 
Type-I 2HDM with the $Z_H$ boson where the $U(1)_H$ boson is fermiophobic. 
This is the simplest solution to the $U(1)_H$ assignments to the SM chiral fermions listed in 
Table I.  Then, $Z_H$ boson can couple with the SM fermions only through the mixing between  
the $\hat{Z}$ and $\hat{Z}_H$ bosons.  In general, the 2HDM with the $Z_H$ boson allows 
wider region compared with the 2HDM  with $Z_2$ symmetry, but if the mixing between 
two Higgs doublets and singlet fields are ignored, there is no essential distinction in the allowed 
regions from the 2HDM with $Z_2$  symmetry.
In particular, if the signal strengths turn out to be close to the SM prediction,  
the distinction would be nontrivial from the Higgs search alone.
Direct search for extra $U(1)_H$ gauge boson and/or extra neutral scalar would be 
important in such a case. 
\item 
In either case, for a given $\mu^{\gamma\gamma}$, the allowed regions for 
$\mu^{WW}$ and $\mu^{ZZ}$ are broader than the ordinary 2HDMs.  
And $\mu^{\tau\tau}$ in Case I could  be smaller than those predicted in the ordinary 2HDMs,
but is similar in Case II. 
On the other hand, it would be difficult to distinguish the ordinary Type-I 2HDM from the model 
with local $U(1)_H$ gauge symmetry based on the observed 126 GeV Higgs signal strengths 
alone, if the data are close to the SM predictions. It would be essential to discover the extra 
scalar bosons and the new gauge boson $Z_H$ in order to tell one from the other. 
\end{itemize}
In this work, we considered only the type-I 2HDMs with $U(1)_H$ gauge symmetry, 
which are the simplest since they are anomaly-free without any extra fermions as long 
as we choose suitable $U(1)_H$ charges for the SM chiral fermions as in Table I.
In this anomaly-free case without extra fermions, it is difficult to enhance 
the signal strengths  $\mu_{gg}^{\gamma\gamma}$ for example.  
On the other hand, more general 2HDMs with $U(1)_H$ gauge symmetry  would  
generically have gauge-anomaly, like in $U(1)_B$ or $U(1)_L$ models.
This gauge anomaly can be cured by adding extra chiral fermions and/or vector-like 
fermions, which would contribute to the production and the decay of Higgs boson
via extra colored and/or electrically charged new particles in the loop and thus   
could enhance $\mu_{gg}^{\gamma\gamma}$.   
It is straightforward to extend the present  analysis to other type of 2HDMs 
with $U(1)_H$ gauge symmetry discussed in Ref.~\cite{Ko-2HDM}, in particular, 
Type-II 2HDM.    These models would have richer structures and be more interesting 
in theoretical and phenomenological aspects, and we plan to report the phenomenological 
analysis on such models in future publications ~\cite{progress}. 

%---------------------------------------------------------------------------
\acknowledgments

We thank Korea Institute for Advanced Study for providing computing resources 
(KIAS  Center for Advanced Computation Abacus System) for this work.
This work was supported in part by Basic Science Research Program through the
National Research Foundation of Korea (NRF) funded by the Ministry of Education Science
and Technology 2011-0022996 (CY), by NRF Research Grant 2012R1A2A1A01006053 
(PK and CY), and by SRC program of NRF funded by MEST (20120001176)
through Korea Neutrino Research Center at Seoul National University (PK).  
The work of PK was also supported in part by Simons Foundation and National 
Science Foundation under Grant No. PHYS-1066293 and the hospitality of the Aspen 
Center for Physics. PK would like to thank CETUP* (Center for Theoretical Underground Physics and Related Areas), supported by the US Department of Energy under Grant No. DE-SC0010137 and by the US National Science Foundation under Grant No. PHY-1342611, for its hospitality and partial support during the 2013 Summer Program.
The work of YO is financially supported by the ERC Advanced Grant project 
“FLAVOUR” (267104).
%---------------------------------------------------------------------------

\appendix

\section{Mass Matrix of CP-even scalars}
\label{sec:Mass Matrix of CP-even scalars}

The mass matrix for $3$ CP-even scalars, $M^2_h$, is
%\begin{equation}
%\begin{eqnarray}
%\begin{pmatrix} M'^2 & M'^2_1 & M'^2_2 \\ M'^2_1 & M^2_{11} & M^2_{12} \\ M'^2_2
%& M^2_{12} & M^2_{22} \end{pmatrix} = \begin{pmatrix}  1 & 0 & 0 \\ 0 & \cos \beta & \sin \beta  \\ 0 &-\sin \beta & \cos \beta   \end{pmatrix}M^2_h \begin{pmatrix}  1 & 0 & 0 \\ 0 & \cos \beta & -\sin \beta  \\ 0 &\sin \beta & \cos \beta   \end{pmatrix}, 
%\end{equation}
\begin{equation}
\begin{pmatrix} 
M'^2 & M'^2_1 & M'^2_2 \\ M'^2_1 & M^2_{11} & M^2_{12} \\ M'^2_2
& M^2_{12} & M^2_{22} \end{pmatrix} 
= \begin{pmatrix}  1 & 0 & 0 \\ 0 & \cos \beta & \sin \beta  \\ 0 &-\sin \beta & \cos \beta   \end{pmatrix}M^2_h \begin{pmatrix}  1 & 0 & 0 \\ 0 & \cos \beta & -\sin \beta  \\ 0 &\sin \beta & \cos \beta   
\end{pmatrix}, 
\end{equation}

\begin{eqnarray}
M'^2 &=& \left ( \frac{m'^2_3}{v_{\Phi} \sqrt{2}} -\frac{m''^2_3}{2} \right )v^2 \cos \beta \sin \beta +\lambda_{\Phi} v_{\Phi}^2, \\
M'^2_1 &=& \widetilde{\lambda}_1 v_{\Phi} v \cos^2 \beta + \widetilde{\lambda}_2 v_{\Phi} v \sin^2 \beta - \frac{m'^2_3}{\sqrt{2}} v \sin 2 \beta, \\
M'^2_2 &=& (-\widetilde{\lambda}_1 v_{\Phi} v  + \widetilde{\lambda}_2 v_{\Phi} v) \cos \beta \sin \beta - \frac{m'^2_3}{\sqrt{2}} v \cos 2 \beta, \\
M^2_{11} &=& \lambda_1 v^2 \cos^4 \beta + \lambda_2 v^2 \sin^4 \beta + (\lambda_3+\lambda_4) \frac{v^2}{2} \sin^2 2 \beta, \\
M^2_{22} &=& \frac{m^2_3}{\cos \beta \sin \beta}+ (\lambda_1+ \lambda_2 )v^2 \cos^2 \beta \sin^2 \beta - (\lambda_3+\lambda_4) \frac{v^2}{2} \sin^2 2 \beta, \nonumber \\
&&  \\
M^2_{12} &=&-(\lambda_1  \cos^2 \beta - \lambda_2 \sin^2 \beta )\frac{v^2}{2} \sin 2 \beta  + (\lambda_3+\lambda_4) \frac{v^2}{2} \sin 2 \beta \cos 2 \beta,  \nonumber \\
&&
\end{eqnarray}
with $m'^2_3(v_{\Phi})\equiv \pa_{\Phi}m^2_3(v_{\Phi}) $ and  $m''^2_3(v_{\Phi})\equiv \pa^2_{\Phi}m^2_3(v_{\Phi}) $.

When $M'^2_1=M'^2_2=0$ is satisfied, the following relations are satisfied:
\begin{eqnarray}
m_H^2&=& M^2_{11} \cos^2 (\alpha - \beta) +M^2_{22} \sin^2 (\alpha - \beta) + M^2_{12} \sin 2 (\alpha - \beta), \\
m_h^2&=& M^2_{11} \sin^2 (\alpha - \beta) +M^2_{22} \cos^2 (\alpha - \beta) - M^2_{12} \sin 2 (\alpha - \beta), \label{eq:CP-even} \\
\tan 2 (\alpha -\beta) &=& \frac{2 M^2_{12}}{M^2_{11}-M^2_{22}}, \label{eq:angle}  \\
m_h^2 +m_H^2-m_A^2&=&  \lambda_1 v^2 \cos^2 \beta + \lambda_2 v^2 \sin^2 \beta. \label{eq:relation-2HDM}
\end{eqnarray}

When $\alpha_{1,2}$ are small,
the angles are approximately
\begin{eqnarray}
\alpha_{1}&=& \frac{-M'^2_1 \sin (\alpha-\beta)+M'^2_2 \cos (\alpha-\beta)}{M'^2- m_h^2} +O((\alpha_{1,2})^2), \\
\alpha_{2}&=& \frac{M'^2_1 \cos (\alpha-\beta)+M'^2_2 \sin (\alpha-\beta)}{M'^2- m_H^2} +O((\alpha_{1,2})^2).
\end{eqnarray}

\end{document}